\documentclass[12pt]{article}
\usepackage{epsfig,amsfonts,amsthm}
\usepackage{color}
\usepackage{amsmath,amssymb}
\usepackage{array}
\usepackage{cancel}
\usepackage{slashed}
\usepackage{cite}

\newcommand{\Et}{\cancel{E}_{T}}
\newcommand{\GeV}{{\ensuremath\rm \,GeV}}

\newcommand{\be}{\begin{equation}}
\newcommand{\ee}{\end{equation}}
\newcommand{\bea}{\begin{eqnarray}}
\newcommand{\eea}{\end{eqnarray}}

\newcommand{\doublet}[2]{ \left( \begin{array}{c}#1 \\ #2 \end{array}\right) }

\def\lsim{\mathrel{\rlap{\lower4pt\hbox{\hskip1pt$\sim$}}
    \raise1pt\hbox{$<$}}}         
\def\gsim{\mathrel{\rlap{\lower4pt\hbox{\hskip1pt$\sim$}}
    \raise1pt\hbox{$>$}}}         

\usepackage{amssymb}
\usepackage{subfig}
\usepackage{wrapfig}
\usepackage{graphicx}

\def\<{\left\langle}
\def\>{\right\rangle}

\newcommand{\bt}{\begin{tabular}}
\newcommand{\et}{\end{tabular}}

\usepackage{slashed}
\usepackage{amssymb} 
\usepackage{float}

\usepackage{tikz}
\usetikzlibrary{decorations.pathmorphing,decorations.markings}

\usepackage{graphicx}

\tikzset{
photon/.style={decorate, decoration={snake,amplitude=2pt, segment length=5pt}, draw=black},
particle/.style={draw=black, postaction={decorate}, decoration={markings,mark=at position .5 with {\arrow[draw=black]{>}}}},
antiparticle/.style={draw=black, postaction={decorate}, decoration={markings,mark=at position .5 with {\arrowreversed[draw=black]{>}}}},
gluon/.style={decorate, draw=black, decoration={coil,amplitude=4pt, segment length=5pt}},
goldstone/.style={draw=green,postaction={decorate},decoration={markings,mark=at position .5 with {\arrow[draw=blue]{>}}}}
}

\begin{document}

\title{\vspace{-20mm}
\begin{flushright}
\normalsize{DIAS-STP-23-12}
\end{flushright}
\hfill ~\\[-5mm]
\textbf{\Large A smoking gun signature of the 3HDM
}        
}

\author{\\[-10mm]
\hspace*{-0.6cm}{\normalsize
A. ~Dey$^{1}$\footnote{E-mail: {\tt atri@stp.dias.ie}},\
V. ~Keus$^{1}$\footnote{E-mail: {\tt venus@stp.dias.ie}},\ 
S.~Moretti$^{2,3,4}$\footnote{E-mail: {\tt s.moretti@soton.ac.uk, stefano.moretti@physics.uu.se}},\
C.~Shepherd-Themistocleous$^{4}$\footnote{E-mail: {\tt claire.shepherd@stfc.ac.uk}}}
\\[0.15cm]
\emph{\small $^1$Dublin Institute for Advanced Studies, School of Theoretical Physics} \\
\emph{\small 10 Burlington road, Dublin, D04 C932, Ireland}\\
\emph{\small $^2$School of Physics and Astronomy, University of Southampton,}\\
\emph{\small Southampton, SO17 1BJ, United Kingdom}\\
\emph{\small $^3$Department of Physics and Astronomy, Uppsala University,}\\
\emph{\small Box 516, SE-751 20 Uppsala, Sweden}\\
\emph{\small  $^4$Particle Physics Department, Rutherford Appleton Laboratory,}\\
\emph{\small Chilton, Didcot, Oxon OX11 0QX, United Kingdom}\\[4mm]
 }
\maketitle

\vspace*{-0.75cm}

\begin{abstract}
\noindent
{\footnotesize
We analyse new signals of a 3-Higgs Doublet Model (3HDM) at the Large Hadron Collider (LHC) where only one doublet acquires a Vacuum Expectation Value (VEV),
preserving a $Z_2$ parity. 
The other two doublets are \textit{inert} and do not develop a VEV, leading to a {\it dark scalar sector} controlled by $Z_2$, with 
the lightest CP-even dark scalar $H_1$ being the DM candidate. 
This leads to the loop induced decay of the 
next-to-lightest scalar, $H_2 \to H_1  \ell \bar \ell $ ($\ell =e,\mu$),
mediated by both  dark  CP-odd neutral and charged scalars.
This is a smoking-gun signal of the 3HDM since it is not allowed in the 2-Higgs Doublet Model
(2HDM) with one inert doublet 
and is expected to be important when $H_2$ and $H_1$ are close in mass.
In practice, this signature can be observed
in the cascade decay of the 
SM-like Higgs boson, $h\to H_1 H_2\to H_1 H_1 \ell \bar \ell$
into two DM particles and di-leptons or $h\to H_2 H_2\to H_1 H_1 \ell \bar \ell \ell \bar \ell$
into two DM particles and four-leptons,
where $h$ is produced from gluon-gluon Fusion (ggF). 
In order to test the feasibility of these channels at the LHC,
we devise some benchmarks, compliant with collider, DM and cosmological data, for which the interplay between these production and decay modes is discussed. In particular, we show that the resulting detector signatures, $\Et\; \ell \bar \ell$ or $\Et\; \ell \bar \ell \ell \bar \ell$, with the invariant mass of $ \ell \bar \ell$ pairs much smaller than $m_Z$, can potentially be extracted already from combining Run 2 and 3 data. } 
 \end{abstract}
\thispagestyle{empty}
\vfill
\newpage
\setcounter{page}{1}

\newpage
\section{Introduction}

Nature appears to have chosen the Higgs mechanism for Electro-Weak Symmetry Breaking (EWSB) as the method to confer mass upon fermions and weak gauge bosons. In its most basic form, through a single Higgs {\it doublet}, this mechanism implies the existence of a single Higgs boson, as discovered in 2012 at the Large Hadron Collider (LHC). This minimal Standard Model (SM) aligns harmoniously with numerous experimental findings. Nevertheless, unresolved questions persist, notably concerning the origin of flavor, characterized by the existence of three families of quarks and leptons, as well as the enigma of Dark Matter (DM). These challenges underscore the necessity for an extension Beyond the SM (BSM).

In the light of the coexistence of three distinct families of quarks and leptons, it is plausible to consider the existence three Higgs doublets. This conceptualization draws parallels with the fermionic sector, wherein the replication of particle families is not dictated by the SM gauge group. 
It is thus conceivable that the symmetries governing the three families of quarks and leptons may mirror those dictating the three Higgs doublets. Within such theoretical frameworks, a family symmetry may undergo spontaneous breaking, along with the Electro-Weak (EW) one. However, a residual subgroup could survive, thereby serving as a stabilizing symmetry for a potential (scalar) Dark Matter (DM) candidate.

In specific cases characterized by particular symmetries, it becomes feasible to identify a Vacuum Expectation Value (VEV) alignment that respects the original symmetry of the potential that  would responsible for the stability of the DM candidate.
Within the framework of 3-Higgs-Doublet Models (3HDMs), among the various symmetries that may govern them \cite{Ivanov:2011ae,Ivanov:2012fp,Ivanov:2012hc,Keus:2013hya}, a straightforward option is the introduction of a
$Z_2$ parity, referred to here as `Higgs parity',
which can prevent Flavor Changing Neutral Currents (FCNCs) and possible charge breaking vacua, rendering it an attractive choice in the pursuit of a viable BSM scenario for the Higgs sector.

The following study focuses on the phenomenological analysis of  these 3HDMs, namely, the one
in which the third scalar doublet is even and the first and second inert\footnote{A doublet is termed ``inert'', or at times ``dark" or simply ``scalar'',  since it does not develop a
VEV, nor does it couple to fermions, so as to distinguish it from one which develops a VEV, i.e., an ``active'' Higgs doublet.} doublets are odd
under the $Z_2$ parity.
Here, we assume a vacuum alignment in the 3HDM space of the kind $(0,0,v)$,  which preserves the aforementioned $Z_2$ symmetry (i.e., the Higgs parity).
Thus we are led to consider a model with two inert doublets plus one Higgs doublet (the so called `I(2+1)HDM'). This construct may be regarded as an extension of the model with one inert doublet plus one Higgs doublet\footnote{This model is known in the literature as the Inert Doublet Model (IDM), herein, we refer to it as I(1+1)HDM, thus, again clarifying the number of inert and active Higgs doublets.} proposed in 1976 \cite{Deshpande:1977rw} and studied extensively for the last few years (see, e.g., \cite{Ma:2006km,Barbieri:2006dq,LopezHonorez:2006gr}).  The lightest neutral scalar (or pseudoscalar) field amongst the two inert doublets,
which are odd under the $Z_2$ parity, provides a 
viable DM candidate which is stabilised by the conserved $Z_2$ symmetry, displaying phenomenological characteristics notably different from the candidate emerging from the I(1+1)HDM case \cite{Grzadkowski:2010au}, both in the CP-Conserving (CPC) and CP-Violating (CPV) cases, as noted in Refs.~\cite{Keus:2014jha,Keus:2014isa,Keus:2015xya,Cordero-Cid:2016krd,Cordero:2017owj,Cordero-Cid:2018man,Keus:2019szx,Cordero-Cid:2020yba,Keus:2020ooy}. Within this framework, we study some new SM-like Higgs decay channels offered by the extra inert fields, with the intent of isolating those which would enable one to distinguish between the I(2+1)HDM and I(1+1)HDM, assuming a CPC scenario throughout. 

The layout of the paper is as follows. In the next section, we introduce the CPC I(1+1)HDM and dwell on the theoretical and
experimental constraints acting on its parameter space. We then discuss the dynamics of the aforementioned inert cascade decays and present a cut-based Monte Carlo (MC) analysis extracting these at the LHC. We finally summarize and conclude.

\section{The CPC I(2+1)HDM}
\label{3HDM}

\subsection{The potential with a $Z_2$ symmetry }

In a model with several Higgs doublets, the scalar potential which is symmetric under a group $G$ of phase rotations can be written as the sum of $V_0$, the phase invariant part, and $V_G$, a collection of extra terms ensuring the symmetry group $G$~\cite{Ivanov:2011ae}.

In this paper, we study a 3HDM symmetric under a $Z_2$ symmetry with the generator
\be
\label{generator} 
g_{Z_2}=  \mathrm{diag}\left(-1, -1, +1 \right), 
\ee
where the doublets, $\phi_1,\phi_2$ and $\phi_3$, have odd, odd and even $Z_2$ quantum numbers, respectively. Note that the vacuum alignment $(0,0,v)$ respects this symmetry.
To ensure the absence of significant FCNCs in the model, the fermions are assumed to be even under the $Z_2$-symmetry and hence only couple to the active scalar doublet, $\phi_3$.
The potential symmetric under the $Z_2$ symmetry in Eq.~\eqref{generator} can be written as 
\bea
\label{V-3HDM}
V  &=& V_0 + V_{Z_2}, \\
V_0 &=& - \mu^2_{1} (\phi_1^\dagger \phi_1) -\mu^2_2 (\phi_2^\dagger \phi_2) - \mu^2_3(\phi_3^\dagger \phi_3) 
\nonumber\\
&&+ \lambda_{11} (\phi_1^\dagger \phi_1)^2+ \lambda_{22} (\phi_2^\dagger \phi_2)^2  + \lambda_{33} (\phi_3^\dagger \phi_3)^2 \\
&& + \lambda_{12}  (\phi_1^\dagger \phi_1)(\phi_2^\dagger \phi_2)
 + \lambda_{23}  (\phi_2^\dagger \phi_2)(\phi_3^\dagger \phi_3) + \lambda_{31} (\phi_3^\dagger \phi_3)(\phi_1^\dagger \phi_1) \nonumber\\
&& + \lambda'_{12} (\phi_1^\dagger \phi_2)(\phi_2^\dagger \phi_1) 
 + \lambda'_{23} (\phi_2^\dagger \phi_3)(\phi_3^\dagger \phi_2) + \lambda'_{31} (\phi_3^\dagger \phi_1)(\phi_1^\dagger \phi_3),  \nonumber\\
V_{Z_2} &=& -\mu^2_{12}(\phi_1^\dagger\phi_2)+  \lambda_{1}(\phi_1^\dagger\phi_2)^2 + \lambda_2(\phi_2^\dagger\phi_3)^2 + \lambda_3(\phi_3^\dagger\phi_1)^2  + {\rm h.c.} 
\eea
This potential has only a $Z_2$ symmetry and no larger accidental symmetry\footnote{Note that adding extra $Z_2$-respecting terms, $(\phi_3^\dagger\phi_1)(\phi_2^\dagger\phi_3)$,  $(\phi_1^\dagger\phi_2)(\phi_3^\dagger\phi_3)$,  $(\phi_1^\dagger\phi_2)(\phi_1^\dagger\phi_1)$,  $(\phi_1^\dagger\phi_2)(\phi_2^\dagger\phi_2)$,  
does not change the phenomenology of the model. The coefficients of these terms, therefore, have been set to zero for simplicity. }.

We shall not consider CP violation in this paper, therefore we require all parameters of the potential to be real.
The full Lagrangian of the model is as follows:
\be 
{ \cal L}={ \cal L}^{\rm SM}_{ gf } +{ \cal L}_{\rm scalar} + {\cal L}_Y(\psi_f,\phi_{3}) \,, \quad { \cal L}_{\rm scalar}=T-V\, ,
\label{lagrbas}
\ee
where ${\cal L}^{\rm SM}_{gf}$ contains the boson-fermion interactions as in the SM, ${ \cal L}_{\rm scalar}$ describes the scalar 
sector including the kinetic term which
has the standard form of $ T = \sum_i \left(D_{\mu} \phi_{i}\right)^{\dagger} \left( D^{\mu} \phi_{i} \right)$ with $D^\mu$ being the covariant derivative for an $SU(2)$ doublet, and $V$ the potential in Eq.~\eqref{V-3HDM}. The ${\cal L}_Y(\psi_f,\phi_{3})$ term describes the Yukawa interaction with $\phi_3$ the only active doublet to play the role of the SM-Higgs doublet.

\subsection{Mass eigenstates}
\label{section-masses}

The minimum of the potential is realized for the following point:
\be 
\phi_1= \doublet{$\begin{scriptsize}$ \phi^+_1 $\end{scriptsize}$}{\frac{H^0_1+iA^0_1}{\sqrt{2}}},\quad 
\phi_2= \doublet{$\begin{scriptsize}$ \phi^+_2 $\end{scriptsize}$}{\frac{H^0_2+iA^0_2}{\sqrt{2}}}, \quad 
\phi_3= \doublet{$\begin{scriptsize}$ G^+ $\end{scriptsize}$}{\frac{v+h+iG^0}{\sqrt{2}}}, 
\label{explicit-fields}
\ee
with 
$
v^2= \frac{\mu^2_3}{\lambda_{33}} .
$

\vspace{5mm}
\noindent The mass spectrum of the scalar particles are as follows.

\begin{itemize}
\item \textbf{The fields from the active doublet $\phi_3$:}\\[2mm]
The $\phi_3$ doublet plays the role of the SM-Higgs doublet with the fields $G^0,G^\pm$ as the would-be Goldsone bosons and $h$as  the SM-like Higgs boson with mass-squared
\be 
m^2_{h}= 2\mu_3^2,
\ee
which has been set to $(125~\GeV)^2$ in our numerical analysis.

\item \textbf{The CP-even neutral fields from the inert doublets $\phi_1$ and $\phi_2$:}\\[2mm]
The inert  CP-even neutral weak basis states, $H^0_{1},H^0_{2}$, are rotated by the angle $\theta_h$ with
\be 
R_{\theta_h}= 
\left( \begin{array}{cc}
\cos \theta_h & \sin \theta_h \\
-\sin \theta_h & \cos \theta_h\\
\end{array} \right), \qquad \mbox{and} \quad \tan 2\theta_h = \frac{2\mu^2_{12}}{\mu^2_1 -\Lambda_{\phi_1} - \mu^2_2 + \Lambda_{\phi_2}},  
\label{diagH}
\ee
into the mass eigenstates, $H_1, H_2$, with squared masses 
\bea
&& m^2_{H_1}=  (-\mu^2_1 + \Lambda_{\phi_1})\cos^2\theta_h + (- \mu^2_2 + \Lambda_{\phi_2}) \sin^2\theta_h -2\mu^2_{12} \sin\theta_h \cos\theta_h, \nonumber\\
&& m^2_{H_2}=  (-\mu^2_1 + \Lambda_{\phi_1})\sin^2\theta_h + (- \mu^2_2 + \Lambda_{\phi_2}) \cos^2\theta_h + 2\mu^2_{12} \sin\theta_h \cos\theta_h, 
\label{eq:CP-even-masses}
\eea
where for a neater notation, we have introduced the $\Lambda_{\phi_i}$ parameters to be
\be 
\Lambda_{\phi_1}= \frac{1}{2}(\lambda_{31} + \lambda'_{31} +  2\lambda_3)v^2, 
\quad \Lambda_{\phi_2}= \frac{1}{2}(\lambda_{23} + \lambda'_{23} +2\lambda_2 )v^2 .
\ee

\item \textbf{The CP-odd neutral fields from the inert doublets $\phi_1$ and $\phi_2$:}\\[2mm]
The inert CP-odd neutral weak basis states, $A^0_{1}, A^0_{2}$, are rotated by the angle $\theta_a$ with
\be 
R_{\theta_a}= 
\left( \begin{array}{cc}
\cos \theta_a & \sin \theta_a \\
-\sin \theta_a & \cos \theta_a\\
\end{array} \right), \qquad \mbox{with} \quad \tan 2\theta_a = \frac{2\mu^2_{12}}{\mu^2_1 - \Lambda''_{\phi_1} - \mu^2_2 + \Lambda''_{\phi_2}},\nonumber
\ee
into the mass eigenstates, $A_1, A_2$, with squared masses 
\bea
&& m^2_{A_1}= (-\mu^2_1 + \Lambda''_{\phi_1})\cos^2\theta_a + (- \mu^2_2 + \Lambda''_{\phi_2}) \sin^2\theta_a -2\mu^2_{12} \sin\theta_a \cos\theta_a, \nonumber\\
&& m^2_{A_2}= (-\mu^2_1 + \Lambda''_{\phi_1})\sin^2\theta_a + (- \mu^2_2 + \Lambda''_{\phi_2}) \cos^2\theta_a + 2\mu^2_{12} \sin\theta_a \cos\theta_a, \nonumber\\
&& \mbox{where} \quad \Lambda''_{\phi_1}= \frac{1}{2}(\lambda_{31} + \lambda'_{31} - 2\lambda_3)v^2 , 
\quad \Lambda''_{\phi_2}= \frac{1}{2}(\lambda_{23} + \lambda'_{23} -2\lambda_2 )v^2.   
\qquad \qquad 
\eea

\noindent
Note that, since the model is CPC, there is no mixing between CP-even and CP-odd states.

\item \textbf{The charged fields from the inert doublets $\phi_1$ and $\phi_2$:}\\[2mm]
The charged Weak basis states, $\phi^\pm_{1}, \phi^\pm_{2}$, are rotated by the angle $\theta_c$ with
\be 
R_{\theta_c}= 
\left( \begin{array}{cc}
\cos \theta_c & \sin \theta_c \\
-\sin \theta_c & \cos \theta_c\\
\end{array} \right), \qquad \mbox{with} \quad \tan 2\theta_c = \frac{2\mu^2_{12}}{\mu^2_1 - \Lambda'_{\phi_1} - \mu^2_2 + \Lambda'_{\phi_2}}, \nonumber
\ee
into the mass eigenstates, $H^\pm_1, H^\pm_2$, with squared masses
\bea
&& m^2_{H^\pm_1}=  (-\mu^2_1 + \Lambda'_{\phi_1})\cos^2\theta_c + (- \mu^2_2 + \Lambda'_{\phi_2}) \sin^2\theta_c -2\mu^2_{12} \sin\theta_c \cos\theta_c, \nonumber\\
&& m^2_{H^\pm_2}= (-\mu^2_1 + \Lambda'_{\phi_1})\sin^2\theta_c + (- \mu^2_2 + \Lambda'_{\phi_2}) \cos^2\theta_c + 2\mu^2_{12} \sin\theta_c \cos\theta_c, \nonumber\\
&& \mbox{where} \quad \Lambda'_{\phi_1}= \frac{1}{2}(\lambda_{31})v^2  , 
\quad \Lambda'_{\phi_2}= \frac{1}{2}(\lambda_{23} )v^2.  
\qquad \qquad \qquad \qquad \qquad \qquad \qquad \quad 
\eea
\end{itemize}

We can categorize the inert particles into two groups, or generations. The second generation is characterized by greater mass compared to the fields in the first generation. Specifically, we will denote the collection of $(H_1, A_1, H^\pm_1)$ as representing the first-generation fields and $(H_2, A_2, H^\pm_2)$ as representing the second-generation fields.

Each of the four neutral particles has the potential to serve as the DM candidate, as long as it is lighter than the other neutral states. To simplify our discussion, we will assume, without loss of generality, that the CP-even neutral particle $H_1$ from the first generation has a lower mass than all other inert particles\footnote{Other neutral scalars could also play the role of DM candidate, e.g., $A_1$ would be the lightest particle after transformation $\lambda_{2,3} \to - \lambda_{2,3}$.  We could also choose $H_2$ to be the lightest particle with $\mu_{12}^2 \to -\mu_{12}^2$, or $A_2$ if both $\lambda_{2,3} \to - \lambda_{2,3}$ and $\mu_{12}^2 \to -\mu_{12}^2$. Hence, the results of our analysis are also applicable to all neutral scalars following suitable sign changes.}, specifically
\be 
m_{H_1} < m_{H_2}, m_{A_{1,2}}, m_{H^\pm_{1,2}}.
\ee

Throughout the remainder of this paper, we will use the notations $H_1$ and DM particle interchangeably, along with their respective properties, such as $m_{H_1}$ and $m_{\rm DM}$.

\subsection{Simplified couplings in the I(2+1)HDM}
\label{simplified}

Because the I(2+1)HDM  involves a large number of free parameters, making it unwieldy to analyse it in its entirety, here, we concentrate on a simplified scenario. That is, the parameters associated with the first inert doublet are scaled by a factor of $n$ compared to those related to the second doublet, as done in \cite{Keus:2014jha,Keus:2014isa,Keus:2015xya,Cordero-Cid:2016krd,Cordero:2017owj,Cordero-Cid:2018man,Keus:2019szx,Cordero-Cid:2020yba,Keus:2020ooy}:
\be 
\label{lambda-assumption} 
\mu^2_1 = n \mu^2_2, \quad \lambda_3 = n \lambda_2, \quad \lambda_{31} = n \lambda_{23}, \quad \lambda_{31}' = n \lambda_{23}', 
\ee
resulting in
\be 
\Lambda_{\phi_1} = n \Lambda_{\phi_2}, \quad \Lambda'_{\phi_1} = n\Lambda'_{\phi_2}, \quad \Lambda''_{\phi_1} = n \Lambda''_{\phi_2},
\ee
without introducing any new symmetry in the potential. 
The parameter $n$ was dubbed as as the \textit{dark hierarchy} parameter in \cite{Cordero-Cid:2018man}.
This simplification is done with the motivation that in the $n=0$ limit the model reduces to the well-known I(1+1)HDM. The $n=1$ limit, or the \textit{dark democracy} limit is studied in detail in \cite{Keus:2014jha, Keus:2015xya,Cordero-Cid:2016krd,Cordero:2017owj}. 
We do not assume any specific relationships among the other parameters of the potential. It is important to emphasize that the remaining quartic parameters, $(\lambda_{1,11,22,12}, \lambda'_{12})$, do not influence the discussed DM phenomenology of the model. Consequently, their values have been set in accordance with the constraints detailed in Sect.~\ref{constraints} and in alignment with the unitarity results presented in \cite{Moretti:2015cwa}.

With this simplification, we can derive analytical expressions for the potential parameters based on selected physical parameters. In this study, our chosen input parameters consist of $(m_{H_1}, m_{H_2}, g_{H_1 H_1 h}, \theta_a, \theta_c, n)$, where $g_{H_1 H_1 h}$ represents the coupling between the SM-like Higgs and the DM candidate. The relevant parameters of the model are subsequently defined as follows:
\bea
&& \Lambda_{\phi_2} = \frac{v^2 g_{H_1 H_1 h}}{4(\sin^2 \theta_h + n \cos^2 \theta_h)},\\
&& \Lambda'_{\phi_2} = \frac{2 \mu_{12}^2}{(1-n) \tan 2 \theta_c}+\mu_2^2,
\\
&& \Lambda''_{\phi_2} = \frac{2 \mu_{12}^2}{(1-n) \tan 2 \theta_a }+\mu_2^2, 
\\
&& \mu_2^2 = \Lambda_{\phi_2} - \frac{m_{H_1}^2+m_{H_2}^2}{1+n},
\\
&& \mu_{12}^2 = \frac{1}{2} \sqrt{(m_{H_1}^2-m_{H_2}^2)^2 - (-1+n)^2 (\Lambda_{\phi_2} - \mu_2^2)^2 },
\\
&& \lambda_2 = \frac{1}{2v^2} (\Lambda_{\phi_2}  - \Lambda''_{\phi_2} ),\\
&& \lambda_{23} = \frac{2}{v^2} \Lambda'_{\phi_2}, 
\\
&& \lambda'_{23} = \frac{1}{v^2} (\Lambda_{\phi_2}  + \Lambda''_{\phi_2} - 2 \Lambda'_{\phi_2} ).
\eea
The mixing angle in the CP-even sector, $\theta_h$, is given by the masses of $H_1$ and $H_2$ and the \textit{dark hierarchy} parameter $n$:
\be
\label{eq:tanthetah}
\tan^2 \theta_h = \frac{m_{H_1}^2 - n m_{H_2}^2}{n m_{H_1}^2 - m_{H_2}^2}.
\ee
Note that we restore the \textit{dark democracy} ($n=1$) limit when $\theta_h = \pi/4$. 
As can be seen from Eq.~\eqref{eq:tanthetah}, for the correct definition of $\tan^2 \theta_h$, the following two relations need to be satisfied: 
$m_{H_1}^2 < n m_{H_2}^2$ and $m_{H_1}^2 < \frac{1}{n} m_{H_2}^2$. 
Without loss of generality, we limit ourselves to $n<1$, which corresponds to $\tan2\theta>0$ for $\theta_h < \pi/4$. Reaching other values of $n$ is a matter of re-parametrising of the potential.

\subsection{Theoretical and experimental constraints}\label{constraints}

The I(2+1)HDM framework is subject to various theoretical, observational and experimental constraints, as discussed in~\cite{Keus:2014jha,Keus:2014isa,Keus:2015xya,Cordero-Cid:2016krd,Cordero:2017owj,Cordero-Cid:2018man,Keus:2019szx,Cordero-Cid:2020yba,Keus:2020ooy}, which we summarize below. All these bounds are satisfied in all our benchmark scenarios to follow.

\begin{enumerate}
\item 
To meet theoretical constraints, it is essential for the potential to bounded from below, and for the Hessian to exhibit positive-definiteness. We apply these constraints by employing the conservative sufficient limits of
\be
\lambda_{ii} > 0, \quad 
\lambda_{ij} + \lambda'_{ij} > -2 \sqrt{\lambda_{ii}\lambda_{jj}}, \quad 
|\lambda_{1,2,3}|< |\lambda_{ii}|, |\lambda_{ij}|, |\lambda'_{ij}| , \quad i\neq j = 1,2,3.
\ee
We take all couplings to be $|\lambda_i| \leq\,4\,\pi$ in accordance with perturbative unitarity limits. 

\item 
Parameterized by the EW oblique parameters $S,T,U$ \cite{Altarelli:1990zd,Peskin:1990zt,Peskin:1991sw,Maksymyk:1993zm}, inert particles $H_i, A_i, H_i^{\pm}$ may introduce important radiative corrections to gauge boson propagators.
We impose a $2\sigma$ agreement with EW Precision Observables (EWPOs) at $95 \%$ Confidence Level (CL) \cite{Baak:2014ora},
\be 
S = 0.05\pm0.11, \quad T = 0.09\pm0.13, \quad U = 0.01\pm0.11.
\ee
Similar to the 2HDM, this condition requires each charged state to be close in mass with a neutral state, in the dark sector \cite{Dolle:2009fn}.

\item 
The contribution of the inert scalars to the total decay width of the EW gauge bosons  constrains the masses of the inert scalars to be \cite{Agashe:2014kda} (for $i,j,k=1,2$)
\be 
\label{eq:gwgz}
m_{H_i, A_j}+m_{H_{k}^\pm}\geq m_{W^\pm}, \quad
\,m_{H_i}+m_{A_j} \geq m_Z, \quad
\,m_{H_{i}^\pm}+m_{H_{j}^\mp} \geq m_Z \, .
\ee

\item  
Non-observation of charged scalars puts a model-independant lower bound on their mass \cite{Lundstrom:2008ai,Cao:2007rm,Pierce:2007ut} and an upper bound on their lifetime \cite{Heisig:2018kfq} to be
\be 
m_{H_{1,2}^\pm}\,\geq\,70\,\GeV, \qquad
\tau_{H_{1,2}^\pm}\,\leq\,10^{-7} \, s \; \Rightarrow \;
\Gamma^\text{tot}_{H_{1,2}^\pm}\,\geq\,6.58\,\times\,10^{-18}\,\GeV,
\ee
to guarantee their decay within the detector. 
In all our benchmark scenarios, the mass of both charged scalars is above 95 GeV and their decay width, primarily to $H^{\pm}_i \to W^\pm H_j$, is of the order of $10^{-1}$ GeV, which is well within limits.

\item
Any model introducing new decay channels for the SM-Higgs boson is constrained by an upper limit on the Higgs total decay width, $\Gamma^h_\text{tot}\,\leq\,9$ MeV \cite{CMS:2018bwq}, and Higgs signal strengths \cite{Khachatryan:2016vau,Aaboud:2018xdt,Sirunyan:2018ouh}. 
In our model, the SM-like Higgs could decay through $h \to S_iS_j$ where $S_iS_j$ are a pair of CP-even ($H_i H_j$) or CP-odd ($A_i A_j$) inert scalars, provided $m_{S_i}+m_{S_j} < m_h$ and $S_{i,j}$ are long-lived enough ($\tau\,\geq\,10^{-7}$ s). 
As a result, $S_{i,j}$  will not decay inside the detector and therefore contribute to the Higgs \textit{invisible} decay with a branching ratio of
\be
\textrm{BR}(h \to S_i S_j) = \frac{\sum_{i,j} \Gamma(h\to S_iS_j)}{\Gamma_h^{\rm SM}+\sum_{i,j} \Gamma(h \to S_iS_j)}, 
\label{inv_all}
\ee
where
\be
\Gamma(h\to S_i S_j)=\frac{g_{h S_i S_j}^{2}v^2}{32\pi m_{h}^3}
\biggl( \left(m_h^2-(m_{S_i}+m_{S_j})^2 \right) \left(m_h^2-(m_{S_i}-m_{S_j})^2 \right)\biggr)^{1/2},
\ee
which sets strong limits on the Higgs-inert couplings. 
Moreover, the partial decay $\Gamma(h\to \gamma\gamma)$ receives contributions from the inert charged scalars. 
The combined ATLAS and CMS Run I results for Higgs to $\gamma\gamma$ signal strength require $\mu_{\gamma \gamma} = 1.14^{+0.38}_{-0.36}$ \cite{Khachatryan:2016vau}. 
In Run II, ATLAS reports $\mu_{\gamma \gamma} = 0.99^{+0.14}_{-0.14}$ \cite{Aaboud:2018xdt}, and CMS reports $\mu_{\gamma \gamma} = 1.18^{+0.17}_{-0.14}$ \cite{Sirunyan:2018ouh} with both of which we are in  $2\sigma$  agreement.


\item 
DM relic density measurements from the Planck experiment \cite{Ade:2015xua},
\be
\label{eq:planck}
\Omega_{DM}\,h^2\,=\,0.1197\,\pm\,0.0022,
\ee
require the relic abundance of the DM candidate to lie within these bounds if it constitutes 100\% of DM  in the universe.
A DM candidate with $\Omega_{\rm DM} h^2 $ smaller than the observed value is allowed; however, an additional DM candidate is needed to complement the missing relic density. Regions of the parameter space corresponding to values of $\Omega_{\rm DM}h^2$ larger than the Planck upper limit are excluded.
We impose a $3\sigma$ agreement with the observation on the relic abundance of our DM candidate, $H_1$.

In this work, we do not focus on the details of DM annihilation (for detailed discussions see  Refs. \cite{Keus:2014jha,Keus:2015xya,Cordero-Cid:2016krd}). However, we require that the DM candidate of the I(2+1)HDM is in agreement with the upper limit from Planck for our Benchmark Points (BPs). If Eq.~\eqref{eq:planck} is exactly satisfied, then $H_1$ provides 100\% of the DM in the Universe. We also consider cases where $H_1$ has a sub-dominant contribution and the missing relic density is to be provided by an extension of the model. This usually happens where mass splittings between $H_1$ and other inert particles are small as is the case in our BP1 and BP2. In these two cases, the coannihilation channels of $H_1 A_i \to Z \to \ell \bar \ell$ are strong and reduce DM relic density to values below the Planck value, even for very small values of Higgs-DM coupling.


\item 
The latest XENON1T results for DM direct detection experiments \cite{Aprile:2018dbl} and FermiLAT results for indirect detection searches \cite{Fermi-LAT:2016uux} do not constrain the model any further. In our benchmark scenarios, the largest direct detection cross section is $\sigma_{DM-N} \approx 10^{-14}\; pb$ and the largest indirect detection cross section is $\langle v\sigma \rangle \approx 10^{-32} ~ cm^3/s$, both of which are well below the limits \cite{Billard:2013qya}.

Note that for the benchmark scenarios in which the DM relic density is below the Planck value, detection limits should be rescaled, leading to the (relic density dependent) limit of:
\be 
\sigma(m_{H_1}) < \sigma^{\rm LUX}(m_{H_1}) \frac{\Omega^{\rm Planck}}{\Omega_{H_1}}.
\ee
We ensure this limit is satisfied for all studied points.

\end{enumerate}

\section{Inert cascade decays} 
\label{cascades}

In the model under investigation, we have one particle, $H_1$, that is entirely stable because it cannot decay into SM particles due to the conservation of the $Z_2$ symmetry, hence, it is a DM candidate. In contrast, all the remaining inert particles, which also possess odd $Z_2$ charge, are assumed to be heavier than the $H_1$, thus making them inherently unstable. The ensuing decays of the heavier inert particles have the potential to yield distinctive experimental signals specifically relevant to the I(2+1)HDM.

Accessing the inert sector can be obtained through interactions with the SM-like Higgs particle $h$ or through interactions with the massive gauge bosons, $Z$ and $W^\pm$. Subsequently, the heavy inert particle decays into $H_1$ along with either on- or off-shell $W^\pm/Z/\gamma$ states. In this model, $h$ can decay into various pairs of inert particles, each resulting in distinct signatures. In this discussion, we will focus on the decays of $h$ into $H_2 H_1$ and $H_2 H_2$. In such case, as mentioned earlier, we will investigate Higgs production at the LHC through the gluon-gluon Fusion (ggF) and  Vector boson fusion (VBF)  processes.

Interesting production and decay patterns can manifest at both tree and loop levels. 
When protons collide, they generate an off-shell gauge boson $Z^*$, which can subsequently yield the $H_1 A_i$ pair (with $i=1,2$). This is followed by the tree-level decay of $A_i$ into $H_1 Z^{(*)}\to H_1 f\bar f$ or (if kinematically allowed) $H_2 Z^{(*)}\to H_2 H_1 f\bar f$ with the subsequent loop decay of $ H_2 \to H_1 \gamma^* \to H_1 f\bar f$ resulting in a $\Et + 4f$ final state\footnote{Note that we use the notation $ H_2 \to H_1 \gamma^* $ in place of $H^2\to H^1 \ell\bar l$, however, all relevant topologies (thus including boxes) are included, see footnote 7.}. 
Another possible process is through the decay of the initial $h$ state into $H_1 H_2\to H_1 H_1 f\bar f$, through the loop decay $H_2 f\bar f$. Further, $h$ could also decay into a pair of $H_2$ particles if kinematically allowed, in which case the process proceeds through $h \to H_2 H_2 \to H_1 H_1 \, f\bar f f'\bar f'$. Here, $f^{(')}$ represent a fermion, either a lepton or a (light) quark, $f^{(')}=\ell, q$.
Thus, for both decay patterns, the outcome is a signature characterized by $\Et + 2f$ or $\Et + 4f$ (potentially accompanied by a resolved forward and/or backward jet in the case of VB or unresolved ones in the case of ggF). This translates into the observation of either a multi-lepton/multi-jet final state, which can typically be captured by the detectors, accompanied by missing transverse energy, $\Et$, caused by the DM pair escaping direct detection. However, hereafter, we neglect considering jet signatures of the above inert cascade decays to avoid excessive QCD background. Therefore, our two targeted final states are  $\Et + \ell\bar\ell$ or $\Et + 2\ell 2\bar\ell$, where$\ell$ represents either electrons ($e$) or muons ($\mu$). In situations where the mass difference $m_{A_i}-m_{H_1}$ or $m_{H_2}-m_{H_1}$ is sufficiently small (around $2m_e$), only the electron-positron signature would emerge, leading to an intriguing Electro-Magnetic (EM) shower.
For the BPs studied here, the $m_{H_2}-m_{H_1}$ mass splitting is larger, namely 5 GeV for BP1 and 10 GeV for the BP2, respectively, resulting also in multi-muon final states. Our trigger choices will eventually dictate which leptonic signature can be pursued experimentally. 

Crucially, here, it is worth emphasising that the loop decay sequence initiated by $h\to H_1 H_2$ is specific to the I(2+1)HDM case, while the tree-level process induced by $A_1\to H_1 Z^{(*)}$ may also apply to the I(1+1)HDM case. (The I(1+1)HDM contains a CP-odd inert state,$ A_1$, but it does not contain $H_2$ or $A_2$.). Furthermore, when the decays are non-resonant, there is no way of separating the two $A_i$ ($i=1,2$) patterns.
In contrast, detecting and observing the decay $h\to H_1 H_2$ (followed by the loop decay $H_2 \to H_1 \ell  \bar\ell  $) would serve as definitive evidence of the I(2+1)HDM.
 
In the upcoming subsections, we discuss both tree- and loop-level decay modes of inert states into the DM candidate and explore the features of the $\Et + 2  \ell$ and $\Et + 4  \ell$ signatures.

\subsection{Tree-level decays of heavy inert states}
\label{sec:tree-level-decays}

CP-odd neutral and charged scalars could undergo tree-level decays into a lighter inert particle accompanied by a real (or virtual) gauge boson $W^{\pm(*)}$ or $Z^{(*)}$. Assuming the mass ordering  $m_{H_{1,2}} < m_{A_{1,2}} < m_{H^\pm_{1,2}}$, the following tree-level decays are possible: 
\be 
A_i \to Z^{(*)} H_j, \quad  H^\pm_i \to W^{\pm(*)} H_j, \quad H^\pm_i \to W^{\pm(*)} A_j, \qquad (i,j=1,2).
\ee 
For clarity, only diagrams involving $H_1$ in the final state are presented in Fig.~\ref{fig:tree-decays}, as seen in diagrams (a) and (b).
The leptonic decays (splittings) of real (virtual) massive gauge bosons yield $\ell \bar{\ell}$ pairs for $Z^{(*)}$ and $\ell \bar{\nu}$ for $W^{\pm(*)}$.
These processes are primarily influenced by the gauge couplings, resulting in relatively small decay widths, typically on the order of $10^{-2}-10^{-4}$ GeV, for the heavy inert particles. However, it is important to note that these decay widths can increase when there is a significant mass difference between $H_1$ and the other particles. It is worth mentioning that, even if all inert particle masses are relatively close (within around 1 GeV), they all still decay within the detector.

\begin{minipage}{1.0\linewidth}
\centering
\begin{figure}[H]
\centering
\hspace{-10mm}
\begin{tikzpicture}[thick,scale=1.0]
\draw (1.5,0) -- node[black,above,xshift=-0.1cm,yshift=0.0cm] {$ $} (1.5,0.03);
\draw[dashed] (0,0) -- node[black,above,xshift=-0.5cm,yshift=0cm] {$A_{1,2}$} (1.5,0);
\draw[decorate,decoration={snake,amplitude=3pt,segment length=10pt}] (1.5,0) -- node[black,above,xshift=0cm,yshift=0cm] {$Z^{(*)}$} (3,1.0);
\draw[dashed](1.5,0) -- node[black,above,yshift=-0.65cm,xshift=-0.2cm]  {$H_1$} (3,-0.75);
\node at (1.5,-1.5) {(a)};
\end{tikzpicture}
\hspace{0.75cm}
\begin{tikzpicture}[thick,scale=1.0]
\draw (1.5,0) -- node[black,above,xshift=-0.1cm,yshift=0.0cm] {$ $} (1.5,0.03);
\draw[dashed] (0,0) -- node[black,above,xshift=-0.5cm,yshift=0cm] {$H^\pm_{1,2}$} (1.5,0);
\draw[decorate,decoration={snake,amplitude=3pt,segment length=10pt}] (1.5,0) -- node[black,above,xshift=-0.2cm,yshift=0cm] {$W^{\pm(*)}$} (3,1.0);
\draw[dashed](1.5,0) -- node[black,above,yshift=-0.65cm,xshift=-0.2cm]  {$H_1$} (3,-0.75);
\node at (1.5,-1.5) {(b)};
\end{tikzpicture}
\hspace{0.75cm}
\begin{tikzpicture}[thick,scale=1.0]
\draw (1.5,0) -- node[black,above,xshift=-0.1cm,yshift=0.0cm] {$ $} (1.5,0.03);
\draw[dashed] (0,0) -- node[black,above,xshift=-0.5cm,yshift=0cm] {$H_{2}$} (1.5,0);
\draw[dashed] (1.5,0) -- node[black,above,xshift=0cm,yshift=0cm] {$h^{(*)}$} (3,1.0);
\draw[dashed](1.5,0) -- node[black,above,yshift=-0.65cm,xshift=-0.2cm]  {$H_1$} (3,-0.75);
\node at (1.5,-1.5) {(c)};
\end{tikzpicture}
\caption{Tree-level decays of heavy inert states to $H_1$ and on/off-shell $Z$, $W^\pm$ and $h$ bosons.}
\label{fig:tree-decays} 
\end{figure}
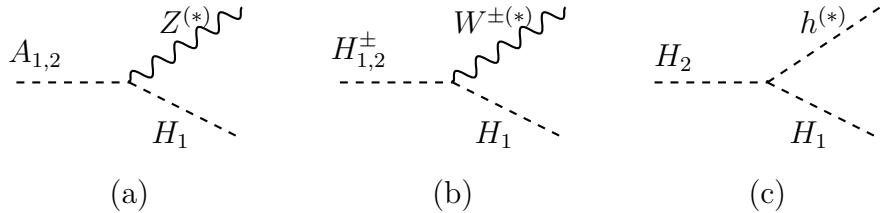    
\end{minipage}\\[1mm]

The heavy CP-even neutral scalar, $H_2$, does not have a coupling to $H_1$ through $Z^{(*)}$, as the CP symmetry remains conserved in our model. Instead, it can decay into the $H_1$ particle along with a Higgs boson, as shown in diagram (c) of Fig.~\ref{fig:tree-decays}. Subsequently, this Higgs particle follows the well-established SM decay patterns. Depending on the mass difference between $H_1$ and $H_2$, the Higgs particle can be significantly off-shell (keeping in mind its SM-like nature necessitates a width of around 4 MeV). Consequently, this results in a relatively narrow decay width for $H_2$ and, in turn, a relatively long lifetime.
However, it is important to clarify that, in all our BPs, this width is not less than $10^{-11}$ GeV, ensuring that the decay of $H_2$ occurs within the detector\footnote{Note that the last diagram in Fig.~\ref{fig:tree-decays} is the one enabling the aforementioned $h\to H_1 H_2$ decay.}. 
As a result, within the scenarios we investigate here, $H_1$ is the sole genuinely invisible dark particle.

\subsection{Loop-level decays of heavy inert states}
\label{sec:loop-decays}

In addition to the previously mentioned tree-level decays, there is also the possibility of loop-mediated processes for a heavy neutral inert particle, $H_2$, as illustrated in Fig.~\ref{fig:radiative}. These processes result in the creation of the lightest inert state, $H_1$ and a virtual photon (scalarly polarised, to preserve spin), which subsequently splits into a pair of light $\ell  \bar\ell  $.\footnote{Detailed calculation of the complete $H_2\to H_1 f \bar{f}$ decays including all topologies can be found in \cite{Cordero:2017owj}.}

\begin{minipage}{1.0\linewidth}
\centering
\begin{figure}[H]
\centering
\begin{tikzpicture}[thick,scale=1.0]
\draw[dashed] (0,0) -- node[black,above,xshift=-1.2cm,yshift=0.0cm] {$H_2$} (2,0);
\draw[dashed] (2,0) -- node[black,above,xshift=0.8cm,yshift=0.0cm] {$H_1$} (4,0);
\draw[photon] (2,0) -- node[black,above,yshift=-0.8cm,xshift=-0.3cm] {$\gamma^*$} (3,-1.5);
\draw[particle](3,-1.5) -- node[black,above,xshift=0.8cm,yshift=0.0cm] {$\ell $} (4,-1);
\draw[antiparticle](3,-1.5) -- node[black,above,xshift=0.8cm,yshift=-0.6cm] {$\bar\ell  $} (4,-2);
\draw[xshift=-0cm] (2,0) node[circle,fill,inner sep=4pt]{} -- (2,0);
\end{tikzpicture}
\caption{Radiative decay of the heavy neutral particle $H_2 \to H_1 \gamma^* \to H_1 \ell  \bar\ell  $.}
\label{fig:radiative}
\end{figure}
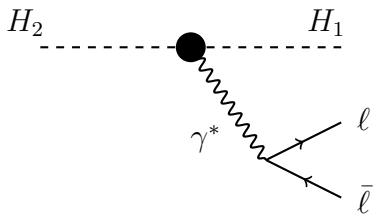
\end{minipage}\\[1mm]

The corresponding loops go through box, triangle and bubble diagrams with $H^\pm_i$ and $W^\pm$ entering, as detailed in Appendix~\ref{app:loop-decay-diagrams}. We may refer to the sum of these one-loop processes as the radiative decay. 
In short, the only (effective) loop-level decay to consider is 
\be
H_2 \to H_1 \gamma^*\,.
\ee
with a BR essentially equal to 1 when the $H_2$ and $H_1$ masses are close.  Notice that, 
in the I(1+1)HDM, there is no counterpart to this process, because CP conservation, which effectively prohibits the only potentially analogous radiative decay within its inert sector (i.e., $A_1\to H_1\gamma^*$). Therefore, as previously mentioned, this signature serves as a means to differentiate between the I(1+1)HDM and models featuring extended inert sectors, such as the I(2+1)HDM.

\subsection{The $\Et \,\ell  \bar\ell  $ and $\Et \, 2\ell  2\bar\ell  $ signatures at the LHC}


In this subsection, we delve into the origins of the distinctive signature discussed earlier, specifically, the missing transverse energy and one and two lepton-antilepton pair(s), $\Et \ell  \bar\ell  $ and $\Et \,2\ell  \,2\bar\ell  $, which can manifest in the I(2+1)HDM. This particular outcome can be generated through both tree-level processes and one-loop decays, as previously elaborated upon. Let's delve deeper into these processes.

The initial mechanism is associated with the decay of the SM-like Higgs boson, which can be produced through the ggF process. Notably, the $hgg$ effective vertex remains identical to that of the SM within the I(2+1)HDM, since the gauge and fermionic sectors of the I(2+1)HDM remain unaltered in comparison to the SM. 
Consequently, the Higgs particle can decay into a pair of neutral CP-odd, CP-even or charged inert particles, here represented as $S_{i,j}$ in Fig.~\ref{fig:Higgsprod}. Depending on the masses of these $S_{i,j}$ particles, they can further undergo decay processes, thereby producing a variety of final states.

\begin{minipage}{1.0\linewidth}
\centering
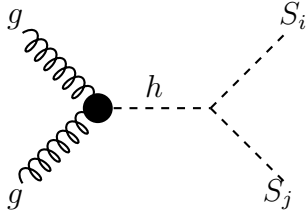
\begin{figure}[H]
\centering
\begin{tikzpicture}[thick,scale=1.0]
\draw[gluon] (0,0) -- node[black,above,xshift=-0.6cm,yshift=0.4cm] {$g$} (1,-1);
\draw[gluon] (0,-2) -- node[black,above,yshift=-1.0cm,xshift=-0.6cm] {$g$} (1,-1);
\draw[dashed] (1,-1) -- node[black,above,xshift=0.0cm,yshift=0.0cm] {$h$} (2.5,-1);
\draw[dashed] (2.5,-1) -- node[black,above,xshift=0.6cm,yshift=0.4cm] {$S_i$} (3.5,0);
\draw[dashed] (2.5,-1) -- node[black,above,yshift=-1cm,xshift=0.4cm] {$S_j$} (3.5,-2);
\draw[xshift=-0cm] (1,-1) node[circle,fill,inner sep=4pt](A){} -- (1,-1);
\end{tikzpicture}
\caption{The ggF-induced production of the SM-like Higgs particle at the LHC with its decay into a pair of inert particles, denoted as $S_i S_j$ which could be $H_i H_j$, $A_i A_j$ or $H^{\pm}_i H^{\pm}_j$ with $i,j=1,2$.}
\label{fig:Higgsprod}
\end{figure}
\end{minipage}\\[2mm]
            
Within the context of the CPC I(2+1)HDM, a process that contributes to the $\Et \ell  \bar\ell  $ signature, representing one of our studied signals, can be expressed as follows:
\be
gg \to h \to H_1 H_2 \to H_1 H_1 \gamma^* \to H_1 H_1 \ell  \bar\ell  ,
\label{first}
\ee
In this process, the off-shell $\gamma^*$ subsequently splits into $\ell  \bar\ell  $, while the $H_1$ states remain undetected.
It is important to note that there exists another tree-level decay of the Higgs boson ($h$) leading to the same signature ($\Et \ell  \bar\ell  $), although not precisely the same final state kinematically. However, these two scenarios are indistinguishable and the process unfolds as follows\footnote{Note that the final state of such process could contain different number of leptons depending on the type and decay channels of intermediate particles. For example, $h \to H^{\pm}_2 H^{\mp}_2 \to H_2 H_2 W^{+(*)} W^{-(*)}$ which, with the subsequent loop decay of $H_2 \to H_1 \gamma^*$, could results in a $\Et + 6\ell$ final state.}:
\be
gg \to h \to H^{\pm}_i H^{\mp}_i \to  H_1  H_1 W^{+(*)} W^{-(*)}\to H_1 H_1 \nu_l \ell  \nu_l \bar\ell    \quad (i=1,2),
\label{second}
\ee 
where the neutrinos escape detection and contribute to the overall missing transverse energy.
In all our benchmark scenarios with the mass ordering $m_{H_{1}}\lesssim m_{H_{2}} < m_{A_{1}} < m_{A_{2}} < m_{H^\pm_{1}} < m_{H^\pm_{2}}$ the contribution of these processes to our signal is sub-dominant.

The process described in Eq.~\eqref{first} is a loop-mediated process which depends on the coupling $g_{H_1 H_2 h}$. Notably, this coupling also affects the relic density of DM. Therefore, in scenarios where this coupling is relatively small, the entire process tends to be suppressed. However, our approach involves maximizing this coupling while ensuring compliance with DM constraints. Furthermore, we adopt a mass spectrum in which the charged Higgs bosons are not excessively heavy, as their high masses would similarly suppress the loop process. In certain parameter configurations, we observe $m_{H_1}+m_{H_2}<m_h$, which results in resonant SM-like Higgs production and loop decay. This resonant behaviour offers a significant enhancement, on the order of $1/\alpha_{\rm EM}$.

In contrast, the process detailed in Eq.~\eqref{second} operates at the tree level, presenting a potential competitive pathway. Nevertheless, for the parameter space explored in our benchmark scenarios, where we aim to maximize the yield of the loop process, this mode becomes practically negligible due to the large charged Higgs masses, precluding the possibility of a resonant interaction with the Higgs boson $h$. 
Similarly, processes in which the SM-like Higgs decays to a pair of CP-odd inert particles\footnote{Similar to the process in Eq.~\eqref{second}, the final state of these processes could contain different number of leptons depending on the type and decay channels of intermediate particles. For example, $h \to A_2 A_2 \to H_2 H_2 Z^{(*)} Z^{(*)}$ which with the subsequent loop decay of $H_2 \to H_1 \gamma^*$ could results in a $\Et + 8\ell$ final state. As mentioned before, in all our benchmark scenarios with the mass ordering $m_{H_{1}}\lesssim m_{H_{2}} < m_{A_{1}} < m_{A_{2}} < m_{H^\pm_{1}} < m_{H^\pm_{2}}$ the contribution of these processes to our signal is sub-dominant.},
\be
gg \to h \to A_i A_j \to  H_1  H_1 Z^{(*)} Z^{(*)}\to H_1 H_1 2\ell  2 \bar\ell    \quad (i=1,2),
\label{eq:AA-decay}
\ee 
are sub-dominant due to the large masses of $A_1$ and $A_2$.
Note that in the above process the off-shell $Z$ could also decay to neutrinos instead of a pair of charged leptons, which clearly changes the final state.

Another process which contributes to the $\Et \, 2\ell  \, 2\bar\ell  $ signature, representing another one of our studied signals, proceeds as follows:
\be
gg \to h \to H_2 H_2 \to  H_1  H_1 \gamma^* \gamma^* \to H_1 H_1 2\ell  2 \bar\ell.
\label{eq:H2H2-decay}
\ee 
In both our BPs, the mass of the $H_2$ pair is below $m_h$, making the Higgs production and loop decay resonant.
As will be shown in Sect.~\ref{sec:collider-analysis}, this signal has very little background, making it the preferred process for our collider analysis.

In principle, there exists another tree-level process that can result in the $\Et \ell  \bar\ell  $ final state within our scenario:
\be 
\label{eq:nstr}
q\bar q \to Z^*\to H_1H_1 Z\to H_1H_1 \ell  \bar\ell  \,.
\ee
This process, illustrated in diagrams (a) and (b) in Fig.~\ref{fig:nstr}, originates from quark-antiquark annihilation and progresses through an $s$-channel off-shell (primary) $Z^*$, where the on-shell (secondary) $Z$ particle ultimately decays into an $\ell  \bar\ell  $ pair. However, there are two significant reasons why we do not prioritize this process.
Firstly, the region of parameter space in which the process described in Eq.~\eqref{first} becomes particularly relevant for LHC phenomenology is where the strength of $g_{H_1 H_2 h}$ is maximal and the Higgs boson $h$ is possibly resonant. This region corresponds to scenarios where the DM relic density is significantly influenced by co-annihilation processes involving $H_1$ and $H_2$\footnote{This is further enhanced when $m_{H_1}\approx m_{H_2}$, which is in fact one of the conditions that we will use in the forthcoming analysis to exalt process in Eq.~\eqref{first} (which is I(2+1)HDM specific) against  the one (also existing in the I(1+1)HDM) that we will be discussing next.}. 
Consequently, this places restrictions on the allowed values of $g_{H_1 H_1 h}$ (especially in the presence of a resonant $h$) coupling. As a result, the process in Eq.\eqref{eq:nstr} becomes less relevant at the LHC.
Secondly, within our framework, the process in Eq.~\eqref{eq:nstr} constitutes a subleading contribution to the invisible Higgs signature of the SM-like Higgs boson (which is primarily dominated by ggF and VBF topologies, extensively investigated in our prior studies in Ref.\cite{Keus:2014isa}). 
In contrast to Eq.~\eqref{first}, this process does not capture any of the heavy scalar states within the model. Thus, it does not provide the means to study the kinematic distributions of the final state able to extract the masses of these heavy scalars by isolating the corresponding thresholds involved in the loops\footnote{In this sense, process in Eq.~\eqref{eq:nstr} would be a background to in Eq.~\eqref{first}, which can be easily removed through a mass veto: $m_{\ell  \bar\ell  }\ne m_Z$.}.
Considering these factors, we will not delve further into the discussion of these two topologies.

\begin{minipage}{0.9\linewidth}
\centering
\begin{figure}[H]
\centering 
\begin{tikzpicture}[thick,scale=1.0]
\hspace{-1cm}
\draw[particle] (0,0) -- node[black,above,xshift=-0.6cm,yshift=0.4cm] {$q_i$} (1,-0.75);
\draw[antiparticle] (0,-1.5) -- node[black,above,yshift=-1.0cm,xshift=-0.6cm] {$\bar{q}_i$} (1,-0.75);
\draw[photon] (1,-0.75) -- node[black,above,xshift=0.0cm,yshift=0.0cm] {$Z^*$} (2,-0.75);
\draw[dashed] (2,-0.75) -- node[black,above,yshift=0.0cm,xshift=-0.0cm] {$h$} (3,-0.75);
\draw[dashed] (3,-0.75) -- node[black,above,yshift=0.1cm,xshift=0.8cm] {$H_1$} (4,-0);
\draw[dashed] (3,-0.75) -- node[black,above,yshift=-0.4cm,xshift=0.8cm] {$H_1$} (4,-1.5);
\draw[photon] (2,-0.75) -- node[black,above,xshift=1.1cm,yshift=-1.5cm] {$Z$} (4,-2.5);
\node at (2,-3) {(a)};
\hspace{1cm}
\draw[particle] (5,0) -- node[black,above,xshift=-0.6cm,yshift=0.4cm] {$q$} (6,-0.75);
\draw[antiparticle] (5,-1.5) -- node[black,above,yshift=-1.0cm,xshift=-0.6cm] {$\bar{q}$} (6,-0.75);
\draw[photon] (6,-0.75) -- node[black,above,xshift=0.0cm,yshift=0.0cm] {$Z^*$} (7,-0.75);
\draw[dashed] (7,-0.75) -- node[black,above,yshift=0.1cm,xshift=1.2cm] {$H_1$} (8.5,-0);
\draw[dashed] (7,-0.75) -- node[black,above,yshift=-0.4cm,xshift=1.1cm] {$H_1$} (8.5,-1.5);
\draw[photon] (7,-0.75) -- node[black,above,xshift=1.15cm,yshift=-1.5cm] {$Z$} (8.5,-2.5);
\node at (6,-3) {(b)};
\hspace{1cm}
\draw[particle] (10,0) -- node[black,above,xshift=-0.6cm,yshift=0.4cm] {$q$} (11,-0.75);
\draw[antiparticle] (10,-1.5) -- node[black,above,yshift=-1.0cm,xshift=-0.6cm] {$\bar{q}$} (11,-0.75);
\draw[photon] (11,-0.75) -- node[black,above,xshift=0.0cm,yshift=0.0cm] {$Z^*$} (12,-0.75);
\draw[dashed] (12,-0.75) -- node[black,above,yshift=0.1cm,xshift=0.8cm] {$H_1$} (13,-0);
\draw[dashed] (12,-0.75) -- node[black,above,yshift=-0.3cm,xshift=-0.3cm] {$A_{1,2}$} (12,-1.75);
\draw[dashed] (12,-1.75) -- node[black,above,yshift=0.1cm,xshift=0.8cm] {$H_1$} (13,-1);
\draw[photon] (12,-1.75) -- node[black,above,xshift=0.6cm,yshift=-0.9cm] {$Z^{(*)}$} (13,-2.5);
\node at (11,-3) {(c)};
\end{tikzpicture}
\vspace{-2mm}
\caption{Diagrams leading to the $\Et \ell  \bar\ell  $ final state via the $H_1H_1 Z^{(*)}$ process.}
\label{fig:nstr}
\end{figure}
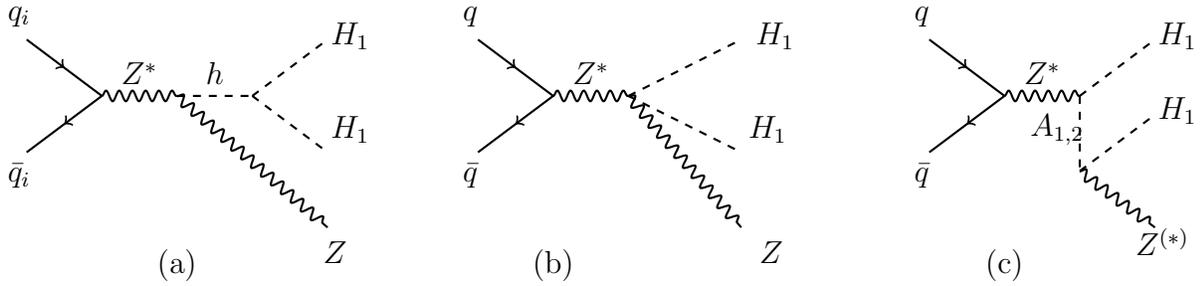
\end{minipage}\\[2mm]

An alternative method to generate the final state $H_1 H_1 \ell  \bar\ell  $ is depicted in graph (c) of Fig.~\ref{fig:nstr}. This process also emerges from $s$-channel quark-antiquark annihilation, producing a virtual neutral massive gauge boson. More explicitly:
\be 
\label{tree}
q\bar q\to Z^* \to H_1 A_i \to H_1 H_1 Z^{(*)}  \to H_1 H_1 \ell  \bar\ell  \qquad(i=1,2),
\ee
wherein the DM candidate is generated alongside a pseudoscalar state and the $Z$ particle may be off-shell. Importantly, this mode proves to be competitive with the one described in Eq.~\eqref{first} within the relevant region of the I(2+1)HDM parameter space.
%
%
Notably, graph (c) in Fig.~\ref{fig:nstr}, in contrast to graphs (a) and (b), due to its heavy pseudoscalar component, may also be isolated in the aforementioned kinematic analysis. {{In fact, in graph (c), the off-shell decay of the $Z$ to leptons pairs can differentiate this process from the those via graphs (a) and (b) in Fig.~\ref{fig:nstr}.}}

{{Furthermore, the processes in such a figure can again produce  $\Et2\ell2\bar\ell$ final states if we replace both or one of the  ${H_1}$'s by $H_2$. In the first case, $Z^{(*)}$ states will decay into pairs of neutrinos contributing to the missing transverse energy. In the second case,  four leptons will naturally emerge from decay of each $H_2 \rightarrow H_1 \gamma^* \rightarrow H_1 \ell  \bar\ell  $. In the first case, the cross section will be suppressed by too many particles in the final state and thus contribute negligibly to the $\Et2\ell2\bar\ell$ final state. In the second case, the relevant process can give a sizeable contribution to it.}}

In conclusion of this subsection, we present a compilation of topologies contributing to VBF production, giving rise to the $\Et \ell  \bar\ell  $ final state. These topologies are illustrated in Fig.~\ref{diag:VBF} and occur before the $H_2\to H_1 \ell  \bar\ell  $ decay. The full production and decay process is thus represented as follows:
\be 
\label{VBF}
q_i q_j \to q_k q_l H_1 H_2 \to H_1 H_1 \gamma^* \to H_1 H_1 \ell  \bar\ell  ,
\ee
where $q_{i,j,k,l}$ symbolizes a(n) (anti)quark of any possible flavor, excluding the top quark. Two notable aspects merit attention in this context. 
Firstly, there is the additional presence of two forward/backward jets, which may or may not be subject to tagging (we will treat these  inclusively). 
Secondly, unlike the case of ggF, not all (gauge-invariant) diagrams correspond to $h\to H_1H_2$ induced topologies (graph (a)), like for  graphs (b) and (c). The primary determinant of which diagram dominates hinges on whether the Higgs boson $h$ can resonate or not. In cases where a resonance is feasible, the first diagram dominates while the last two become competitive only when resonance cannot be attained.
%

\begin{minipage}{0.9\linewidth}
\centering
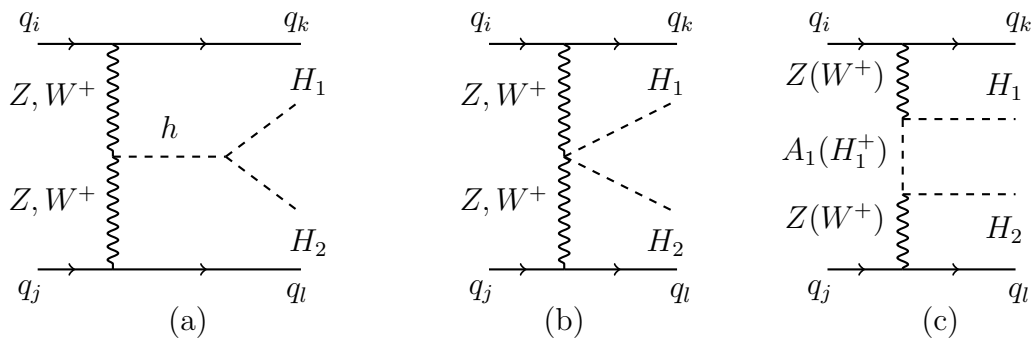
\begin{figure}[H]
\centering       
\hspace{-5mm}   
\begin{tikzpicture}[thick,scale=1.0]
\draw[particle] (0,0) -- node[black,above,xshift=-0.6cm,yshift=0.0cm] {$q_i$} (1,0);
\draw[photon] (1,0) -- node[black,above,xshift=-0.8cm,yshift=-0.3cm] {$Z,W^+$} (1,-1.5);
\draw[photon] (1,-1.5) -- node[black,above,xshift=-0.8cm,yshift=-0.2cm] {$Z,W^+$} (1,-3);
\draw[particle] (0,-3) -- node[black,above,xshift=-0.6cm,yshift=-0.6cm] {$q_j$} (1,-3);
\draw[dashed] (1,-1.5) -- node[black,above,yshift=0.1cm,xshift=-0.0cm] {$h$} (2.5,-1.5);
\draw[dashed] (2.5,-1.5) -- node[black,above,yshift=0.3cm,xshift=0.6cm] {$H_1$} (3.5,-0.75);
\draw[dashed] (2.5,-1.5) -- node[black,above,yshift=-1.1cm,xshift=0.6cm] {$H_2$} (3.5,-2.25);
\draw[particle] (1,0) -- node[black,above,xshift=1.2cm,yshift=0.0cm] {$q_k$} (3.5,0);
\draw[particle] (1,-3) -- node[black,above,xshift=1.2cm,yshift=-0.6cm] {$q_l$} (3.5,-3);
\node at (2,-3.7) {(a)};
\hspace{1cm}
\draw[particle] (5,0) -- node[black,above,xshift=-0.6cm,yshift=0.0cm] {$q_i$} (6,0);
\draw[photon] (6,0) -- node[black,above,xshift=-0.8cm,yshift=-0.3cm] {$Z,W^+$} (6,-1.5);
\draw[photon] (6,-1.5) -- node[black,above,xshift=-0.8cm,yshift=-0.2cm] {$Z,W^+$} (6,-3);
\draw[particle] (5,-3) -- node[black,above,xshift=-0.6cm,yshift=-0.6cm] {$q_j$} (6,-3);
\draw[particle] (6,0) -- node[black,above,xshift=0.8cm,yshift=0.0cm] {$q_k$} (7.5,0);
\draw[particle] (6,-3) -- node[black,above,xshift=0.8cm,yshift=-0.6cm] {$q_l$} (7.5,-3);
\draw[dashed] (6,-1.5) -- node[black,above,yshift=0.3cm,xshift=0.6cm] {$H_1$} (7.5,-0.75);
\draw[dashed] (6,-1.5) -- node[black,above,yshift=-1.1cm,xshift=0.6cm] {$H_2$} (7.5,-2.25);
\node at (6,-3.7) {(b)};
\hspace{1cm}
\draw[particle] (8.5,0) -- node[black,above,xshift=-0.6cm,yshift=0.0cm] {$q_i$} (9.5,0);
\draw[photon] (9.5,0) -- node[black,above,xshift=-0.9cm,yshift=-0.3cm] {$Z (W^+)$} (9.5,-1);
\draw[dashed] (9.5,-1) -- node[black,above,xshift=-0.9cm,yshift=-0.3cm] {$A_1 (H^+_1)$} (9.5,-2);
\draw[photon] (9.5,-2) -- node[black,above,xshift=-0.9cm,yshift=-0.2cm] {$Z (W^+)$} (9.5,-3);
\draw[particle] (8.5,-3) -- node[black,above,xshift=-0.6cm,yshift=-0.6cm] {$q_j$} (9.5,-3);
\draw[particle] (9.5,0) -- node[black,above,xshift=0.8cm,yshift=0.0cm] {$q_k$} (11,0);
\draw[particle] (9.5,-3) -- node[black,above,xshift=0.8cm,yshift=-0.6cm] {$q_l$} (11,-3);
\draw[dashed] (9.5,-1) -- node[black,above,yshift=0.1cm,xshift=0.6cm] {$H_1$} (11,-1);
\draw[dashed] (9.5,-2) -- node[black,above,yshift=-0.8cm,xshift=0.6cm] {$H_2$} (11,-2);
\node at (10,-3.7) {(c)};
\end{tikzpicture}
\vspace{-2mm}
\caption{Diagrams  leading to the $\Et + \ell  \bar\ell  $ final state via VBF topologies.}
\label{diag:VBF}
\end{figure}
\hspace*{3.5mm}
\end{minipage}


{{In summary, despite the presence of several irreducible backgrounds to both our target final states, there are  enough kinematic differences between the various noise and our signals to warrant attempting extracting the latter. Specifically, owing to the small mass difference between $H_2$ and $H_1$ and the fact that the $\gamma^*\to\ell\bar\ell$ splitting in the $H_2\to H_1 2\ell$ decay tends to be soft and collinear (and consequently the dominant contribution, as detailed in the appendix), the invariant mass of same flavor and opposite sign leptons will be much smaller in comparison to the one emerging in both on- and off-shell decays of a $Z$ boson into di-leptons, the exploitation of this feature indeed being the most effective way of disentangling signals and the irreducible backgrounds that we have discussed. (In fact, we shall see that it will also help significantly against non-irreducible ones.)}}



\section{The cut based collider analysis}
\label{sec:collider-analysis}

As discussed earlier, our primary objective is finding discernible signatures which can distinguish between 3HDM and 2HDM scenarios at collider experiments.
We focus on the decay channel $H_2 \rightarrow H_1 \gamma^* \rightarrow H_1 \ell  \bar\ell$ as a smoking gun signature and study its detectability during the high-luminosity phase of the LHC. 
Here, $\ell \equiv \mu$ since muons are well-observed particles in colliders and their misidentification rates are significantly lower compared to those of the electrons.
Moreover, trigger efficiencies for muons have already been increased at the LHC
as will be discussed further. 

Let us emphasize that a signal with missing transverse energy and two leptons in its final state, has a major  drawback; there are substantial and irremovable SM backgrounds to this signal with the main contributions from the $W^\pm W^\mp, \, ZZ$ and $t \bar{t}$ (leptonic) processes, whose cross sections are in the pb range, much larger than the typical cross section in our BPs which is in the fb range\footnote{
The cross section for the $t \bar{t}$ channel is around 300 pb, where both $t$s can decay to bottom quark and $W^\pm$ boson where one or both $W^\pm$ will end up with some leptonic decays and will serve one or two leptons with missing transverse energy in final state. Note that $b$ quarks have a large misidentification rate at the LHC. Jets could also be misidentified as leptons when the $t \bar{t}$ state decays semi-leptonically which substantially reduces our BSM signal significance over the SM background even after applying subsequent cuts.}. 

As a result, we consider a final state with missing transverse energy and at least three muons where our DM candidate, $H_1$ is the dominant source of missing energy. The process under study is $p p \rightarrow H_2 H_2$ where each $H_2$ decays to $H_1 \gamma^* \rightarrow H_1 \,\mu^+ \mu^-$. As discussed before, when the mass difference between $H_2$ and $H_1$ is very small, the BR($H_2 \rightarrow H_1 \gamma^*$) is nearly 100\%. In Tab.~\ref{tab:bps} we present two Benchmark Points (BPs), BP1 and BP2, with $m_{H_2}-m_{H_1}=5$ and 10 GeV, respectively, and in agreement with all constraints discussed in section~\ref{constraints}.

\begin{table}[h!]
\centering
\begin{tabular}{|c|c|c|c|c|c|c|c|c||c|c|}
\hline
Benchmark & $m_{H_1}$ & $m_{H_2}$ & $m_{A_1} $ & $m_{A_2} $  & $m_{H^\pm_1} $ & $m_{H^\pm_2} $
& $n$ &  $\theta_h$  &  $\sigma_{2\mu}$  &     $\sigma_{4\mu}$ \\[1mm]
\hline
BP1 &	50 & 55	& 95 & 104 & 116 & 127 & 0.83 & 0.105 &  0.02224 & 6.923  \\[1mm]
\hline
BP2 &	 50  & 60 & 94	&  112 & 115 & 137  & 0.70 & 0.103 &  0.06  & 4.0 \\[1mm]
\hline
\end{tabular}
\caption{Definition of BPs with the masses shown in GeV. The last two columns show the cross sections for processes $\sigma_{2\mu} \equiv \sigma(pp\rightarrow H_2 H_2 \rightarrow H_1 H_1\, \mu^+ \mu^-)$ and $\sigma_{4\mu} \equiv \sigma(pp\rightarrow H_2 H_2 \rightarrow H_1 H_1\, 2\mu^+  \, 2\mu^-)$ in fb. In both BPs, $\theta_a = 0.03$, $\theta_c = 0.02$ and $g_{hDM}= 0.01$.}
\label{tab:bps}
\end{table}

The cross sections of the events for the signals are computed at leading order with their respective backgrounds computed at the next-to-leading order (NLO) and created using \texttt{Madgraph@MCNLO}~\cite{Alwall:2014hca}. We adopt the \texttt{nn23lo1} parton distribution function. The detector simulation is handled by \texttt{Delphes-3.5.0}~\cite{deFavereau:2013fsa}. We have used the inbuilt detector efficiencies available at \texttt{Delphes-3.5.0} to identify final state isolated muons. We use no further trigger efficiencies and apply \texttt{PYTHIA8}~\cite{Sjostrand:2006za} showering and hadronisation.


\subsection{Signal and backgrounds}
\label{sec:signal-background}

As mentioned before, the process under study is $p p \rightarrow H_2 H_2$ where each $H_2$ decays to $H_1 \gamma^* \rightarrow H_1 \,\mu^+ \mu^-$.
As $H_1$ is the DM candidate, it will escape the detector and only appears as missing transverse energy, $\Et$ in the final state. 
Since the final state muons are the products of the off-shell photon splittings (which invariant mass is $\sim m_{H_2}-m_{H_1}$ and hence very small) they will not be energetic. It is thus a challenging task to identify all muons required, as they will have to survive all detector acceptance cuts and be compliant with some suitable trigger definition. As a result, we require at least three muons in the final sate which can all be tracked in the detector. Subsequently, we identify the signal and background as follows.
\begin{itemize}
\item \textbf{Signal:} 
The signal is
at least three muons with at least one pair of opposite charge muons plus missing transverse energy.

\item \textbf{Background:}
The dominant backgrounds to this signal~\cite{ATLAS:2020fdg,ATLAS:2021wob} are:
\begin{enumerate}
\item 
The di-boson final state $VV$ ($V = W^\pm, \, Z, \, \gamma$) with the largest contributions from the $W^\pm \, Z/\gamma$ and $ZZ$ final state where both vector bosons decay leptonically.

\item 
The tri-boson $VVV$ final state ($V = W^\pm, \, Z, \, \gamma$) where the main contribution is from the $W^\pm \, W^\mp \, Z/\gamma$, $W^\pm W^\mp W^\pm$ and $ZZZ$ final states when all three vector bosons decay leptonically.

\item 
The $t \bar{t} X$ final state ($X = W^\pm, \, Z, \, \gamma, \, W^\pm \, W^\mp, \,t \bar{t}$) where the fully leptonic decay mode leads to at least three leptons with at least one pair of leptons with opposite charge. 
\end{enumerate}
\end{itemize}
Finally, notice that we require minimum two $b$-jets in the final sate, due to the low efficiency of tagging $b$-jets at LHC (which is $\sim 80\%$ at the most and depending on the jet $p_T$).

\subsection{Distributions}
\label{sec:distributions}

In this section, we present an in-depth analysis at the detector level of the distribution patterns of several noteworthy observables for both signal and  backgrounds.
By scrutinising the prospective distribution profiles of these final-state observables, we aim to identify a region exhibiting a more favourable signal-to-background ratio.

\begin{figure}[h!]
\centering
\includegraphics[width=0.495\textwidth]{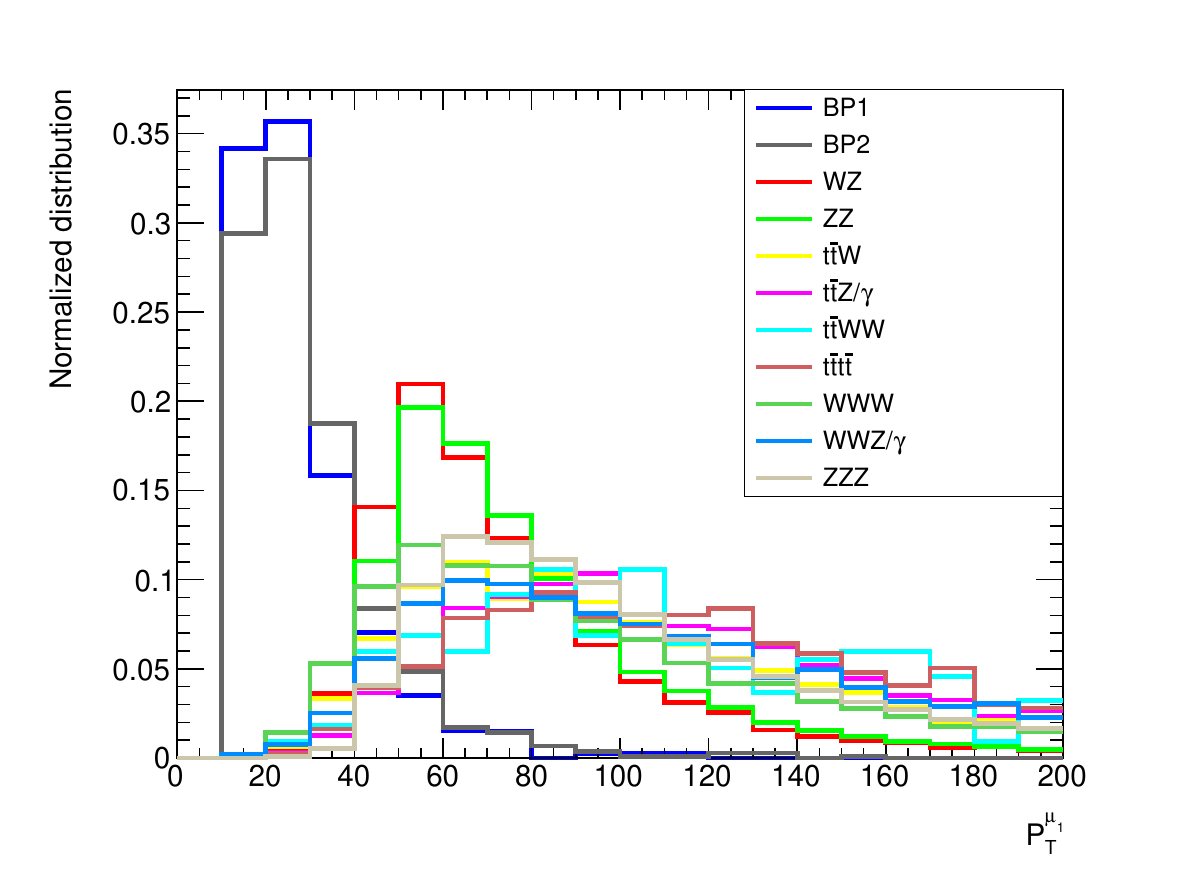}
\includegraphics[width=0.495\textwidth]{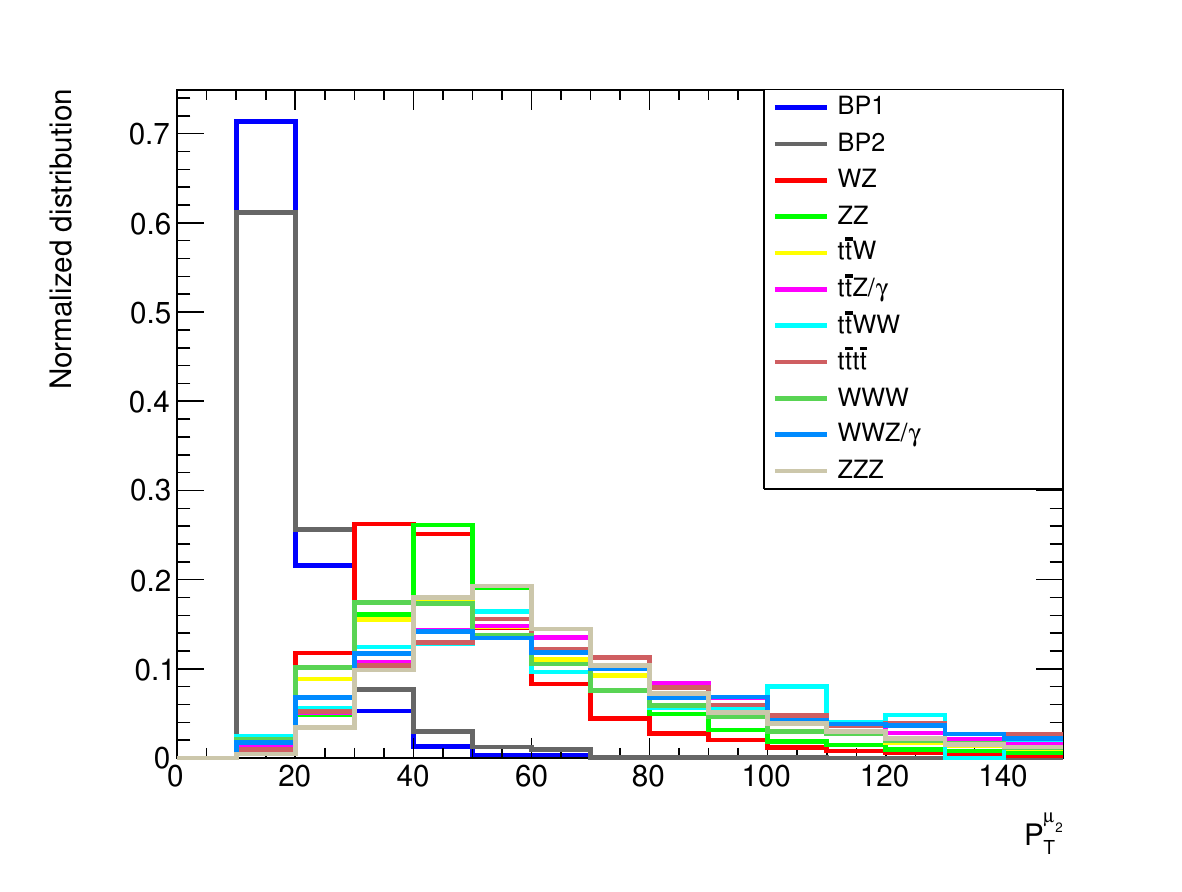} \\
\includegraphics[width=0.495\textwidth]{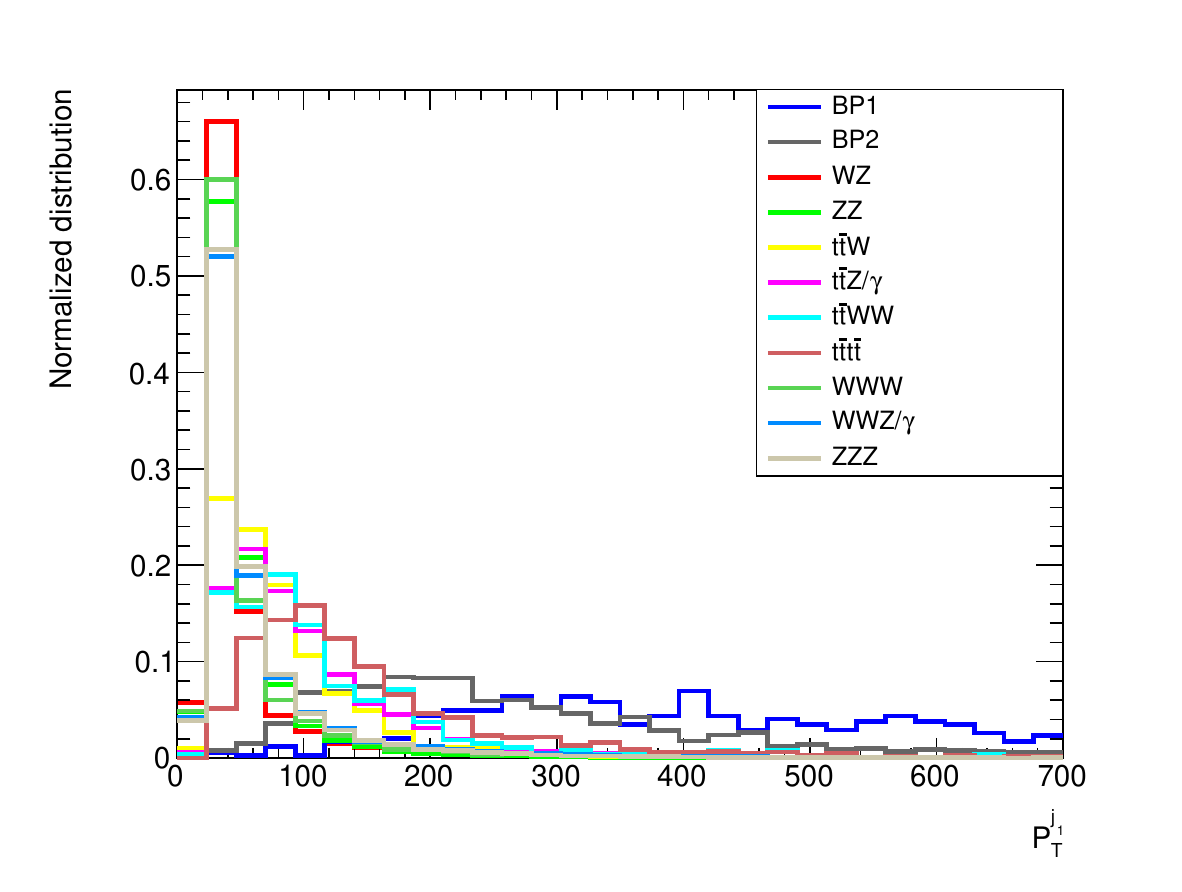}
\includegraphics[width=0.495\textwidth]{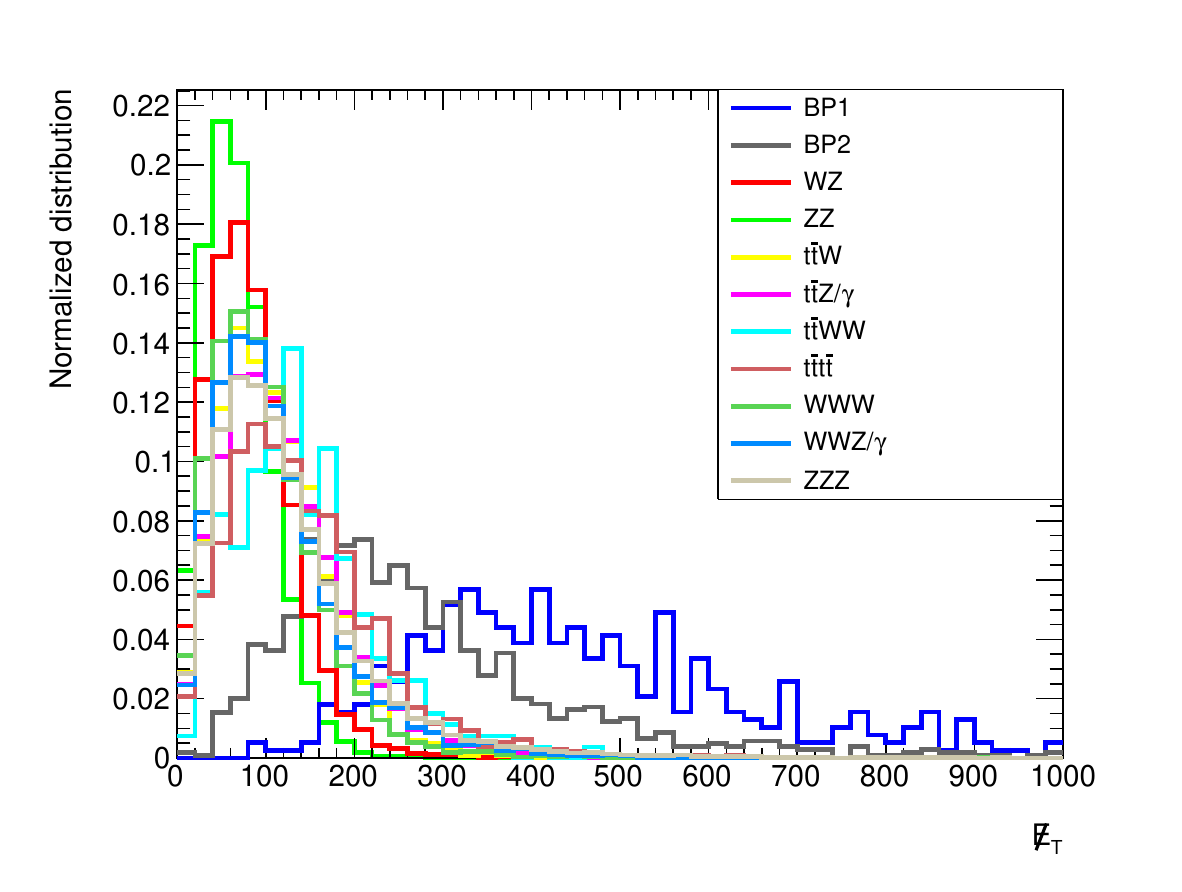}
\caption{Normalised distributions for signal and backgrounds for the $p_T$ of the leading lepton (top left), $p_T$ of the sub-leading lepton (top right), the missing transverse energy $\Et$ (bottom left) and the $p_T$ of the  leading jet (bottom right) after detector analysis.}
\label{fig:dist_1_I_delphes}
\end{figure}

In Fig.~\ref{fig:dist_1_I_delphes}, we show the distribution of the transverse momentum of the leading and sub-leading muons, denoted by $p^{\mu_1}_T$ and $p^{\mu_2}_T$, in the top panel. 
Although the final state of our signal doe not naturally have a hard jet (at tree level), the radiation from the initial state is capable of creating such a jet in the final state. 
Consequently, to provide a comprehensive perspective, we introduce a secondary variable: the transverse momentum of the leading jet, $p^{j_1}_T$, in the lower left panel of Fig.~\ref{fig:dist_1_I_delphes}, alongside the missing transverse energy distribution (in the lower right one). 

Given that the mass difference between $H_2$ and $H_1$ is small, the $p_{T}$ of muons peak at lower values in our BPs, compared to that of the SM backgrounds. As a result, this could be a very useful observable for distinguishing the two. 
In fact, in contrast to the signal, the main contribution to the muons $p_{T}$ in the SM backgrounds is from (on-shell)  $W^\pm$ decay to $\ell \nu$ pairs (which  peaks around $40$ GeV). 
It is also important to note that the source of the missing transverse energy in the signal is the two DM candidates, which can increase the missing transverse energy to higher values when compared to the SM backgrounds.

\begin{figure}[h!]
\centering
\includegraphics[width=0.495\textwidth]{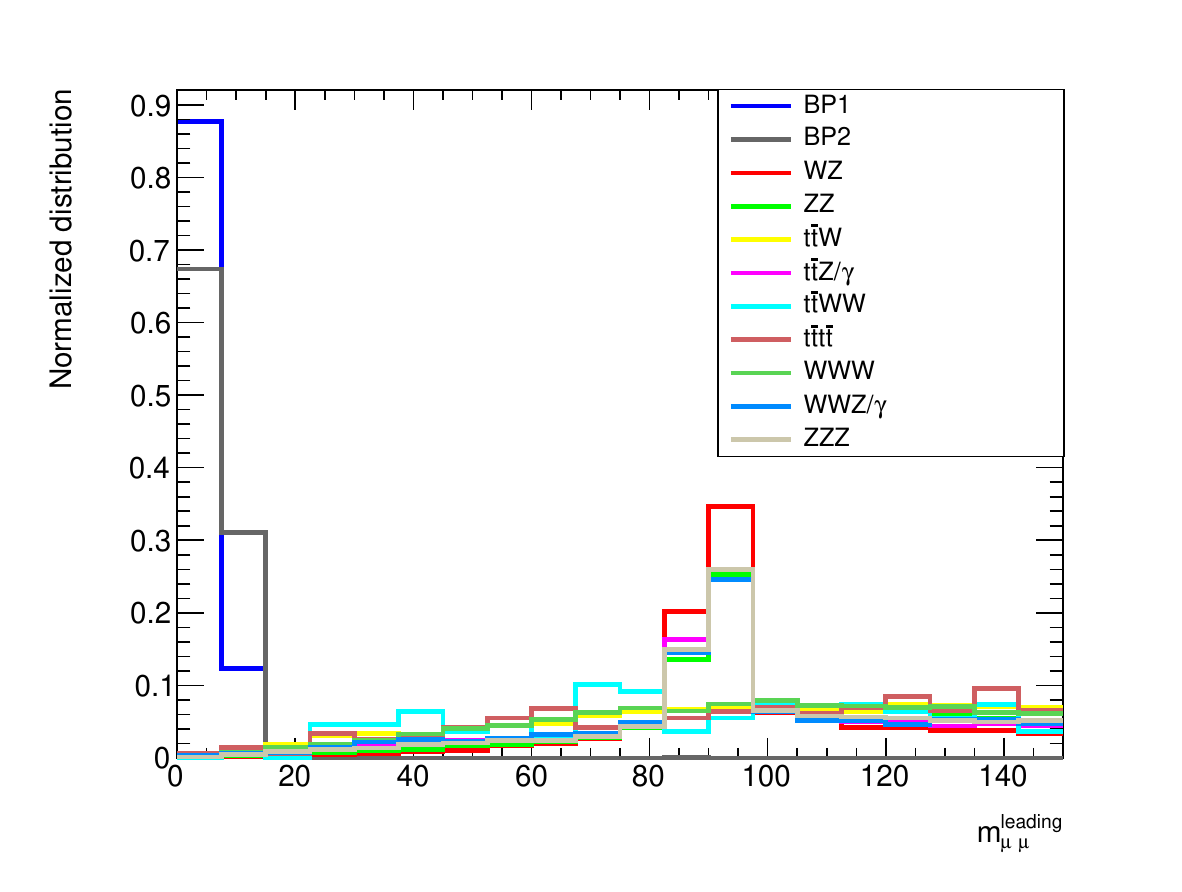}\includegraphics[width=0.495\textwidth]{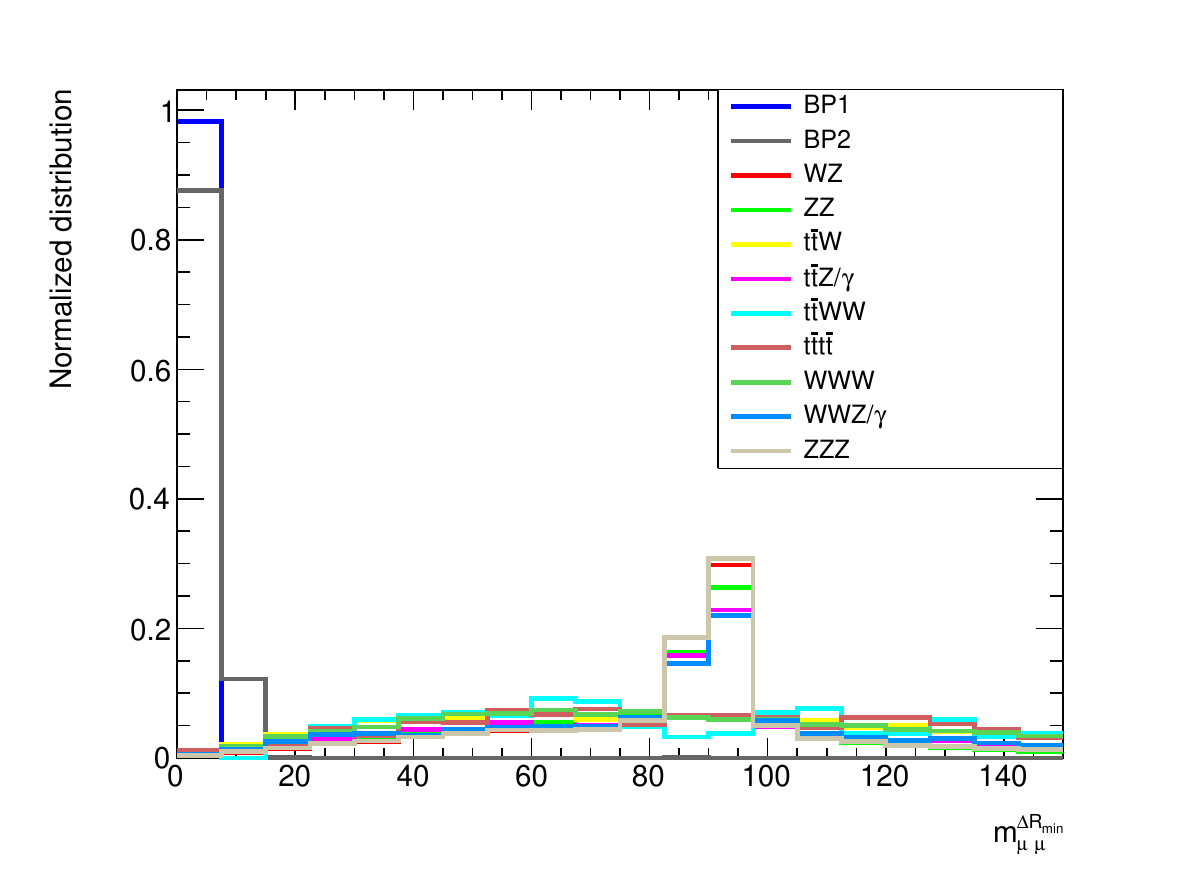} \\
\includegraphics[width=0.495\textwidth]{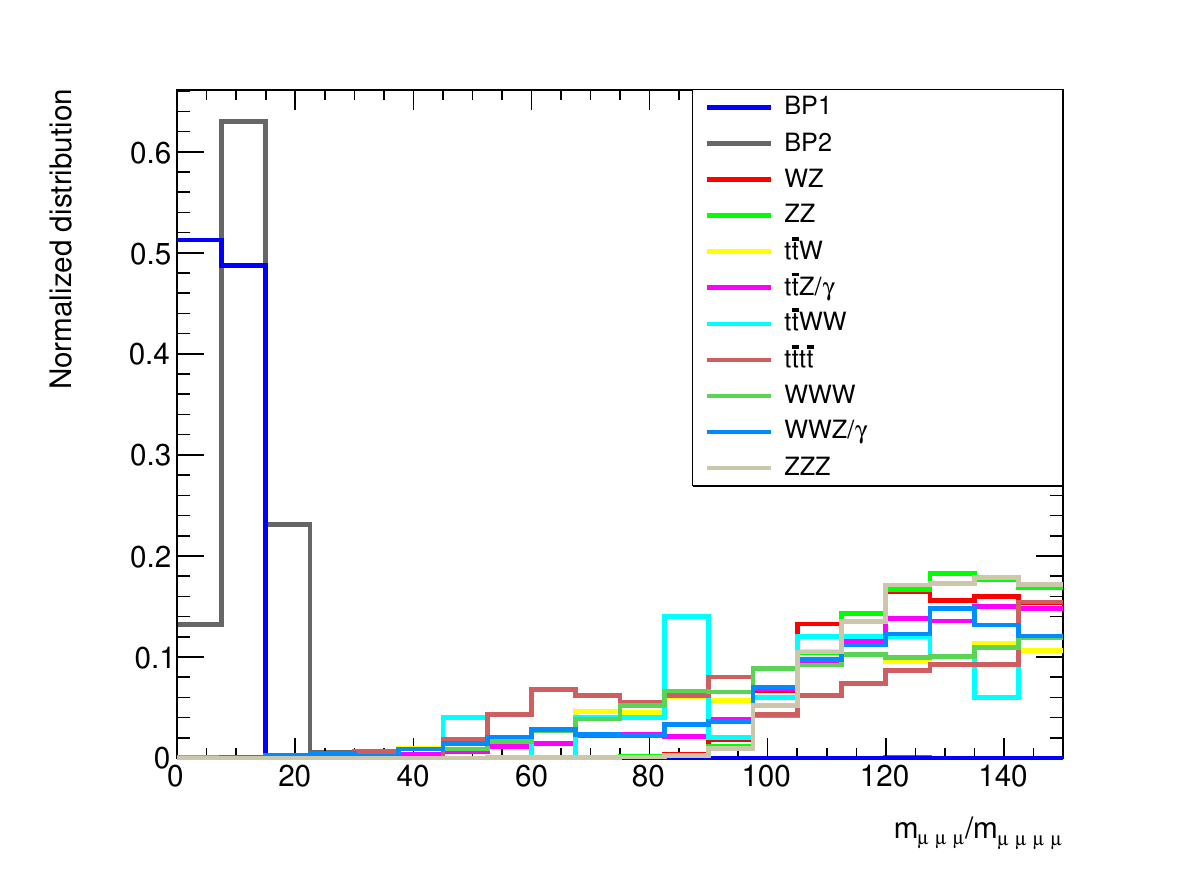}
\caption{Normalised distributions for signal and backgrounds for the invariant mass of the two leading muons (top left), the invariant mass of the two  closest muons one of which being the leading one 
 (top right) and the invariant mass of all muons (bottom) after detector analysis.}
\label{fig:dist_2_I_delphes}
\end{figure}

The analysis of invariant mass distributions is presented in Fig.~\ref{fig:dist_2_I_delphes},  elucidating the fact that these are  key observables. Specifically, our attention is directed towards two sets of such variables in the top panels of Fig.~\ref{fig:dist_2_I_delphes}, namely $m^{\mathrm{leading}}_{ \mu \mu}$ and $m^{\Delta R_{\mathrm{min}}}_{\mu \mu}$. Here, $m^{\mathrm{leading}}_{\mu \mu}$ denotes the invariant mass of the two leading muons while $m^{\Delta R_{\mathrm{min}}}_{\mu \mu}$ represents the invariant mass of the leading muon in conjunction with its closest companion within the angular separation $\Delta R_{\mathrm{min}}$.
In the bottom panel of Fig.~\ref{fig:dist_2_I_delphes}, we show the invariant mass of all final-state muons as a collective variable. Notably, here, one can see that, due to the relatively small $m_{H_2} - m_{H_1}$ mass difference, the invariant masses of leptons tend to peak at lower values whereas they exhibit a tendency to cluster around higher values for the background processes, where they are mostly produced through $W^\pm$ and/or $Z$ decays. 
These variables will assume a pivotal role in the discrimination of our signal from the prevalent SM backgrounds.

\begin{figure}[h!]
\centering
\includegraphics[width=0.495\textwidth]{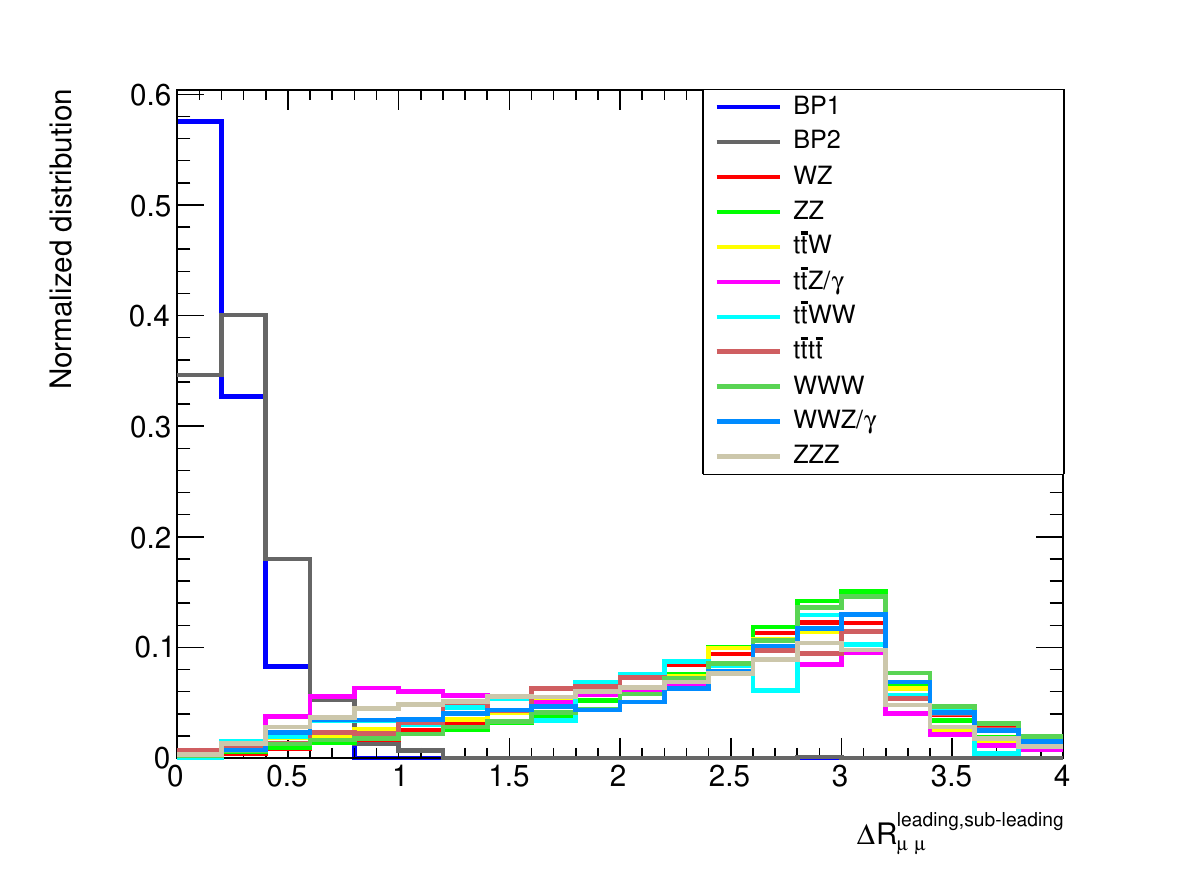}
\includegraphics[width=0.495\textwidth]{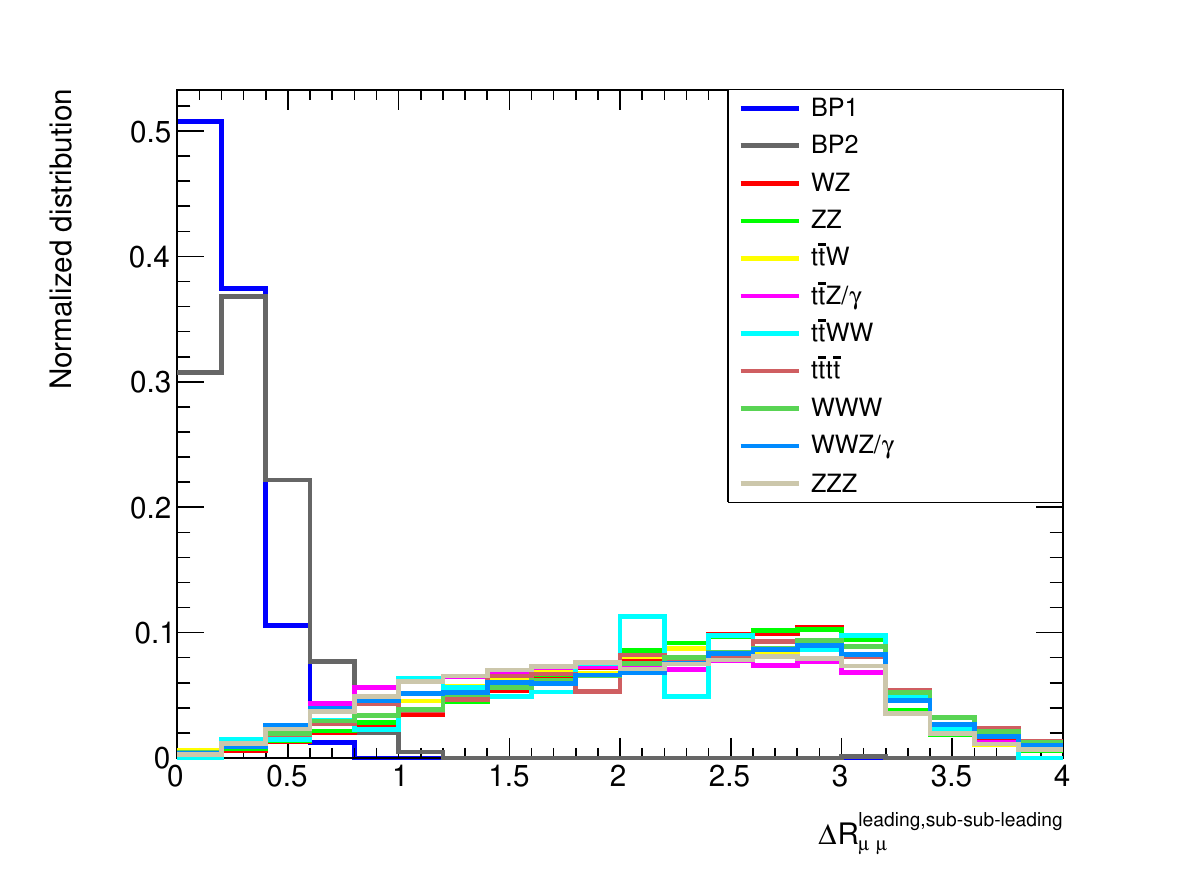} \\
\includegraphics[width=0.495\textwidth]{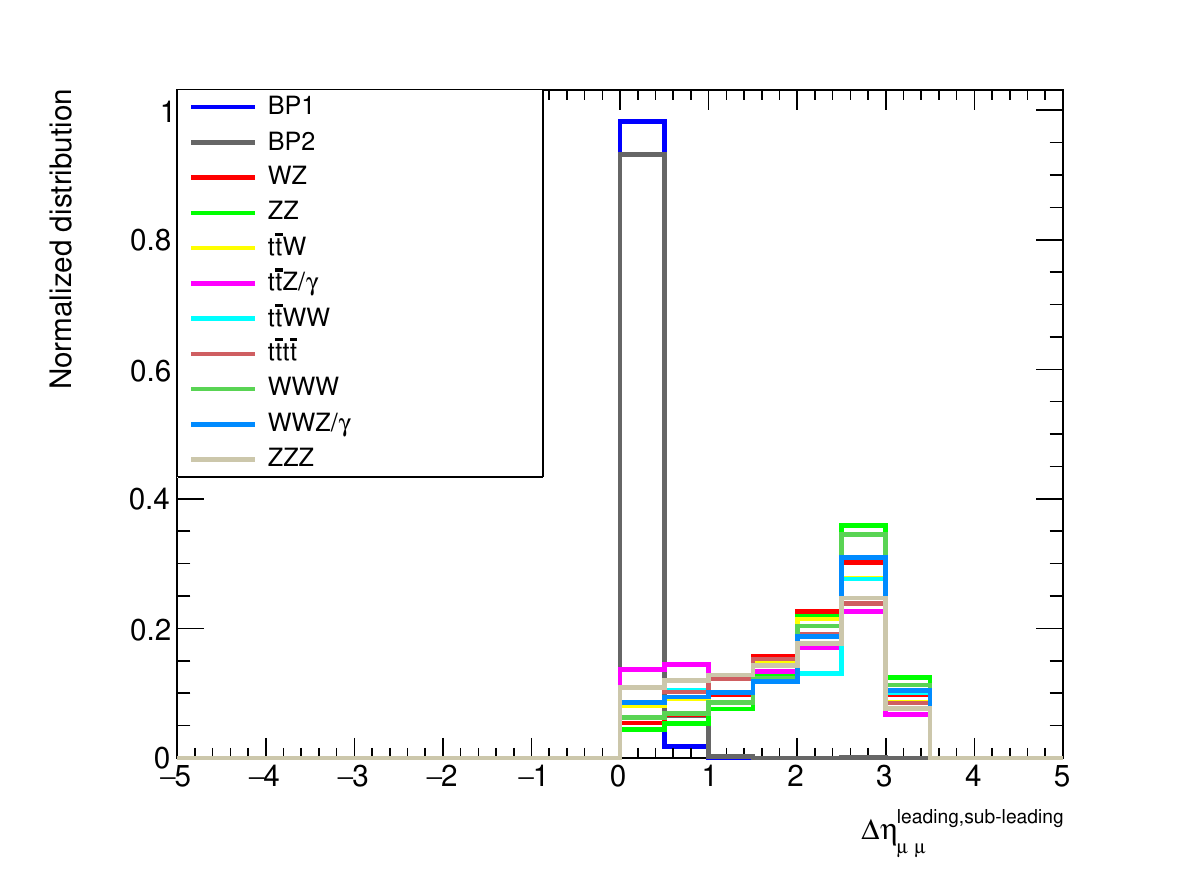}
\includegraphics[width=0.495\textwidth]{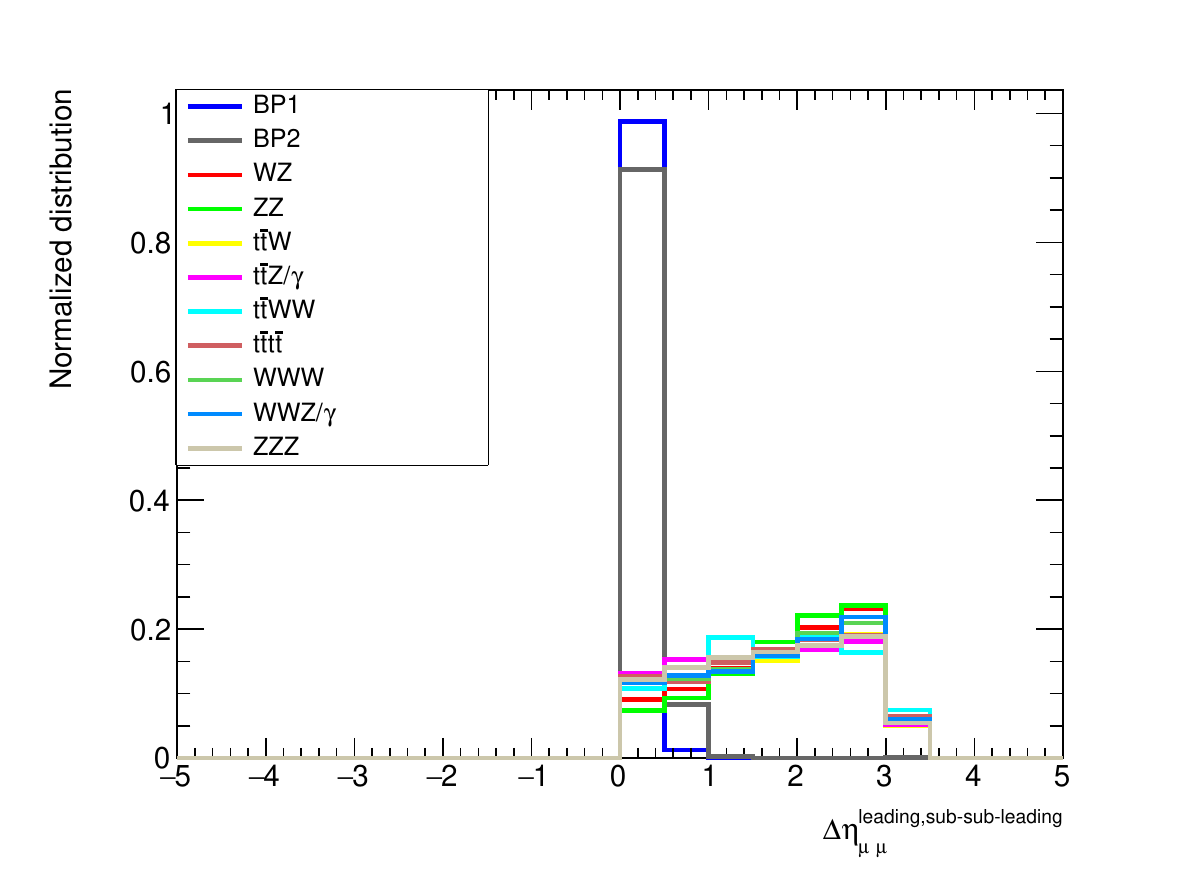}
\caption{Normalised distributions for signal and backgrounds for the $\Delta R$ of the leading and sub-leading muons (top left), the $\Delta R$ of the leading and sub-sub-leading muons (top right), the $\Delta \eta$ of the leading and sub-leading muons (bottom left) and the $\Delta \eta$ of the leading and sub-sub-leading muons (bottom right) after detector analysis.}\label{fig:dist_3_I_delphes}
\end{figure}

In Fig.~\ref{fig:dist_3_I_delphes}, we introduce several variables designed to yield measurements of the spatial separation between the two muons in the final state. In the top panels of Fig.~\ref{fig:dist_3_I_delphes}, we show the radial separation $\Delta R$, as the Euclidean distance in the $(\eta, \phi)$ plane, i.e., $\sqrt{\Delta \eta^2 + \Delta \phi^2}$, of the two leading leptons, leading lepton and sub-sub-leading lepton. 
In the bottom panels of Fig.~\ref{fig:dist_3_I_delphes}, we present the $\Delta \eta$ distribution where $\eta$ denotes pseudo-rapidity, again for the two leading leptons,  leading lepton and sub-sub-leading lepton. Crucially, our analysis reveals a distinctive pattern: the muons in our signal events generally tend to exhibit close proximity while in the case of background events they tend to be more widely separated. To discern between our signal and background events, applying reasonable cuts based on these separation variables will prove to be effective too.

\begin{figure}[h!]
\centering
\includegraphics[width=0.495\textwidth]{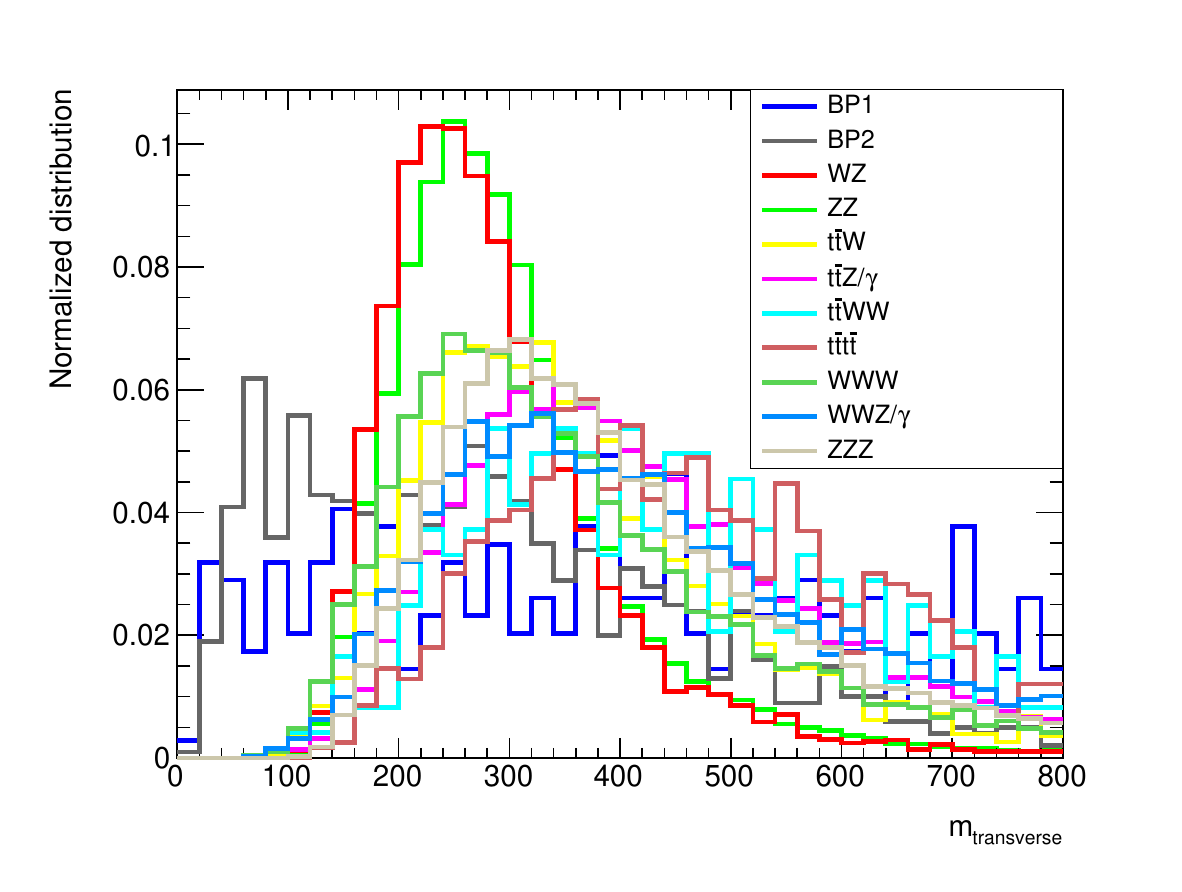}
\includegraphics[width=0.495\textwidth]{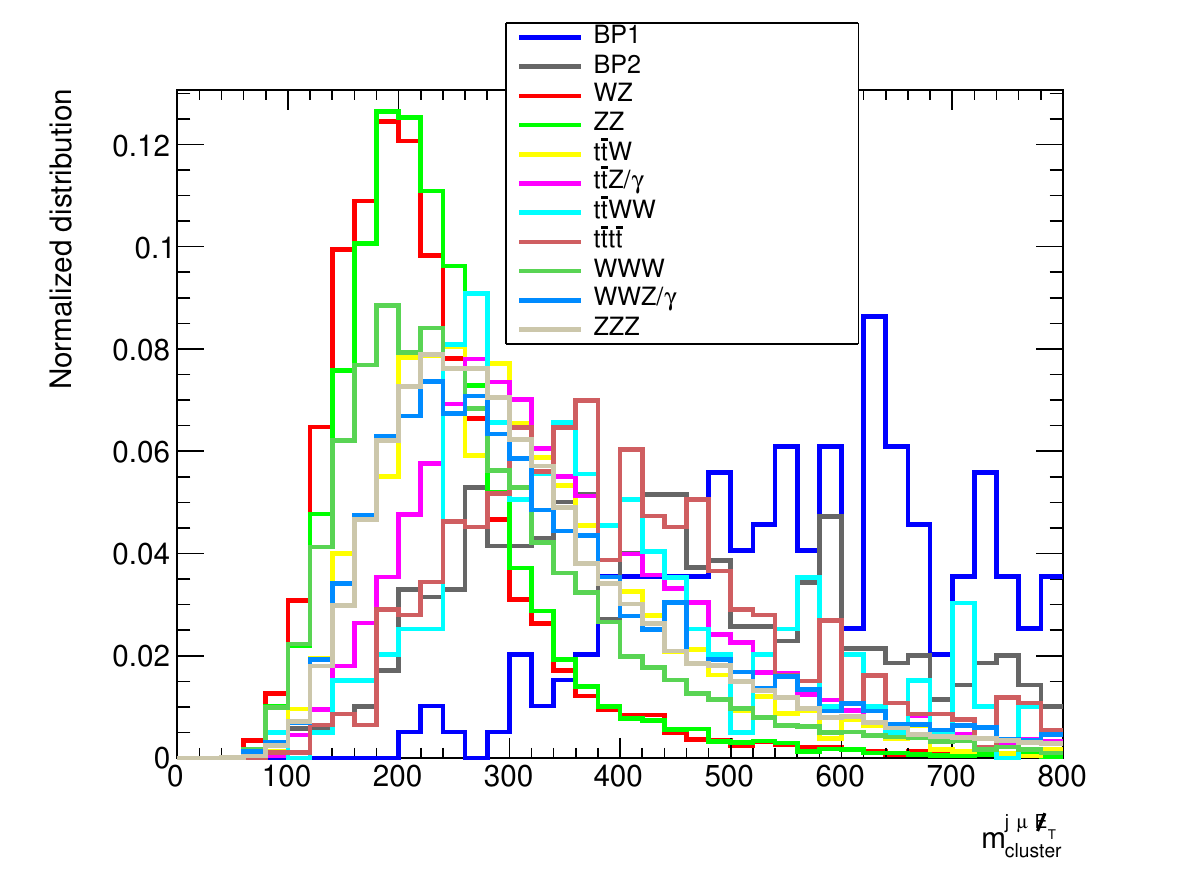}
\caption{Normalised distributions for signal and backgrounds for the transverse mass of all muons $m_{\mathrm{transverse}}$ (left)
and the cluster transverse mass $m_{\mathrm{cluster}}$ (right) after detector analysis.}
\label{fig:dist_4_I_delphes}
\end{figure}


In Fig.~\ref{fig:dist_4_I_delphes}, we show the following variables:
\bea 
\label{eq:m_transverse}
m_{\mathrm{transverse}} &=& \sqrt{\left(\sqrt{\left(\sum_{\mu} p_T \right)^2 + m^2_{\mu \mu \mu/\mu \mu \mu \mu}} + \Et \right)^2-\left(\sum_{\mu} p_T + \Et \right)^2} \, , 
\\[2mm]
m^{j\mu \, \Et}_{\mathrm{cluster}} &=& \sum_{\mu} p_T \, + \, p^{j_1}_{T} + \Et \, ,
\label{eq:m_cluster}
\eea
where $m_{\mathrm{transverse}}$, which represents the transverse mass of the overall muonic final state while  
 $m^{j\mu \, \Et}_{\mathrm{cluster}}$ is defined as the sum of 
the transverse momentum of all detectable muons plus the one of the leading jet  plus the missing transverse energy. 
While the former appears rather useful in separating signal and backgrounds the latter  does not offer a substantial improvement in discriminating the two. 
Altogether, it is evident that such variables do not always provide a very effective means to diminish the backgrounds relatively to the signal, despite the naive expectation that   significant backgrounds such as $W^\pm Z$ and $ZZ$ would tend to accumulate around values approximating the sum of the gauge boson masses, when the mass difference between $H_2$ and $H_1$ is much smaller in comparison. This is due to the fact  that SM-like Higgs state can transfer significant energy to its decay products, owing to boosted behaviour of the partons inside the protons.

\begin{figure}[h!]
\centering
\includegraphics[width=0.495\textwidth]{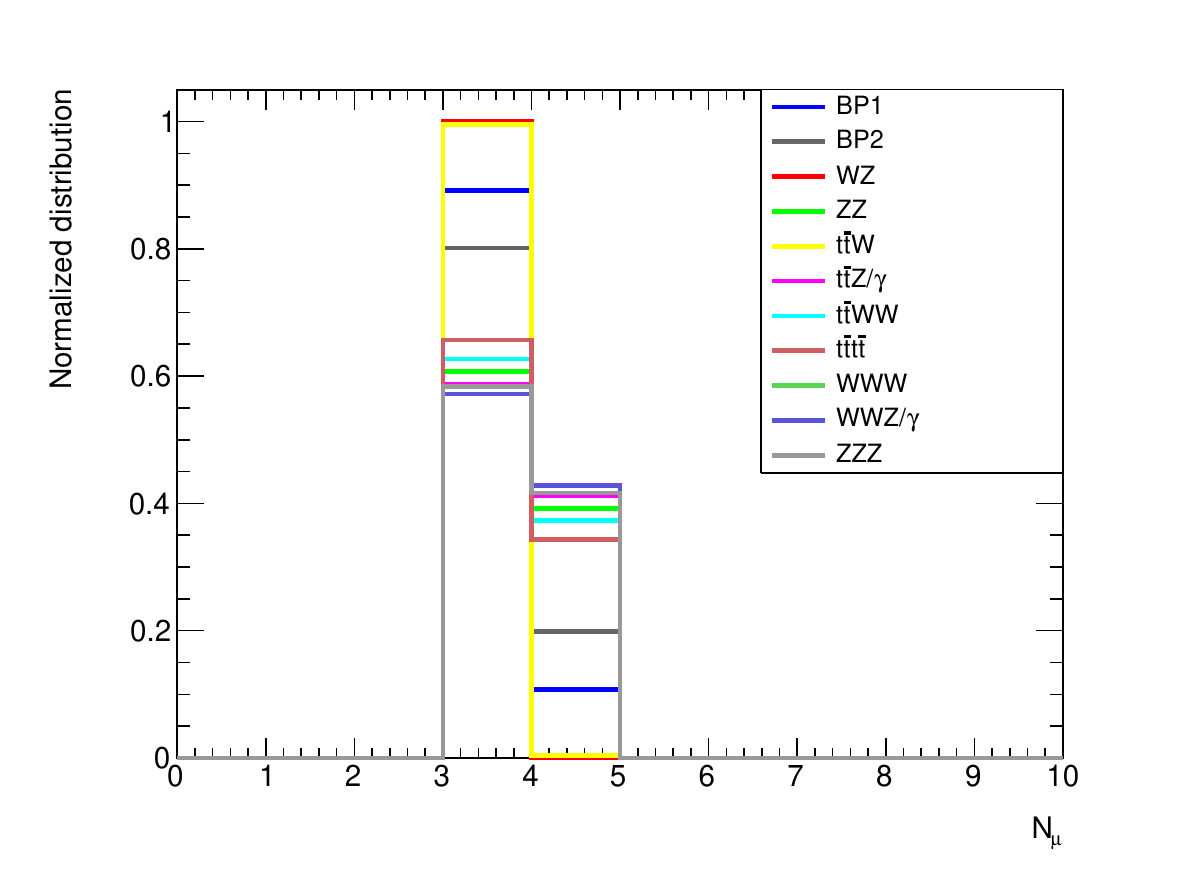}
\includegraphics[width=0.495\textwidth]{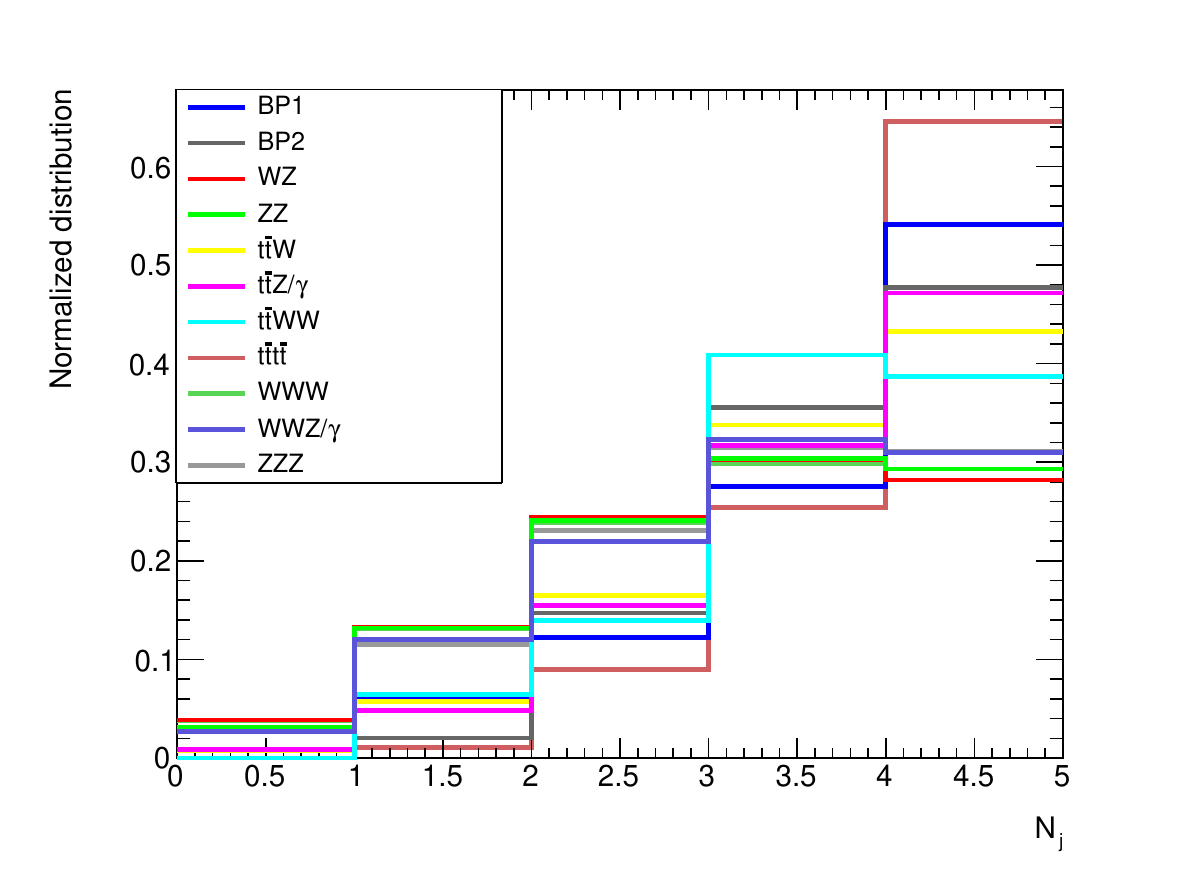} 
\caption{Normalised distributions for signal and backgrounds for the total number of muons $N_{\mu}$ (left) and the total number of jets $N_{j}$ (right) after detector analysis.}
\label{fig:dist_5_I_delphes}
\end{figure}

Finally, in Fig.~\ref{fig:dist_5_I_delphes}, we show the total number of muons, $N_{\mu}$, and the total number of jets, $N_j$, in the final state, neither of which appears particularly useful in separating signal from backgrounds. 
 
\subsection{Selection criteria and results}
\label{Results}

Following the presentation of the various observables in the previous plots, we now proceed to implement regular cuts in order to enhance the signal significance while mitigating the background contributions. Guided by our earlier observations, we apply the following selection criteria to our datasets.

\begin{itemize}
\item 
{\bf Pre-selection cut:}
We select events that feature a final state comprising at least three or four muons, accompanied by missing transverse energy $\Et$, while excluding the presence of any $b$-jets.

\item {\bf Cut-A}: 
We require the following selection criteria:
\begin{enumerate}
\item 
The invariant mass of the two leading muons, denoted by $m^{\mathrm{leading}}_{\mu \mu}$, as well as the invariant mass of the two closest muons  one of which being the leading one, denoted  by $m^{\Delta R{\mathrm{min}}}_{\mu \mu}$, must both fall below a threshold of $50$ GeV.
\item 
The invariant mass calculated for all isolated muons in the final state, denoted by $m_{{\mu \mu \mu}/{\mu \mu \mu \mu}}$, must not exceed $70$ GeV.
\end{enumerate}

\item {\bf Cut-B}: 
The following selection criteria are required:
\begin{enumerate}
\item
The invariant mass of the two leading muons, denoted by $m^{\mathrm{leading}}_{\mu \mu}$, and the invariant mass of the two closes muons, denoted by $m^{\Delta R_{\mathrm{min}}}_{\mu \mu}$, must be below $20$ GeV.
\item
The invariant mass of all final state isolated muons, denoted by $m_{{\mu \mu \mu}/{\mu \mu \mu \mu}}$, must not exceed $30$ GeV.
\item
The radial separation between the two leading muons, denoted as $\Delta R^{\mathrm{leading, sub-leading}}_{\mu \mu}$, remains below $1.0$ and, similarly, the radial separation between the leading muon and the sub-sub-leading muon, denoted by $\Delta R^{\mathrm{leading,sub-sub-leading}}_{\mu \mu}$, should not exceed $1.2$.
\item
The difference in pseudo-rapidity between the two leading muons, $\Delta \eta^{\mathrm{leading, sub-leading}}_{\mu \mu}$, should be less than $1.0$. Similarly, the difference in pseudo-rapidity between the leading muon and the sub-sub-leading muon, denoted by $\Delta \eta^{\mathrm{leading,sub-sub-leading}}_{\mu \mu}$, must also remain below $1.0$.
\end{enumerate}

\item {\bf Cut-C}: 
Subsequent to the application of cut-B, the following selection criteria are enforced:
\begin{enumerate}
\item 
The missing transverse energy, $\Et$, must exceed $100$ GeV.
\item 
The transverse momentum of the leading muon, $p^{\mu_1}_T$, should be less than $40$ GeV.
\item 
The transverse momentum of the sub-leading muon, $p^{\mu_2}_T$, must not exceed $30$ GeV.
\item 
The transverse momentum of the leading jet, $p^{j_1}_T$, is required to be greater than $50$ GeV.
\end{enumerate}

\item {\bf Cut-D}: 
The following selection criteria are imposed:
\begin{enumerate}
\item
The invariant mass of two leading muons, denoted by $m^{\mathrm{leading}}_{\mu \mu}$, and the invariant mass of the two closest muons  one of which being the leading one, denoted by $m^{\Delta R_{\mathrm{min}}}_{\mu \mu}$, must be below $20$ GeV.
\item
The invariant mass of all final state isolated muons, denoted by $m_{{\mu \mu \mu}/{\mu \mu \mu \mu}}$, must not exceed $30$ GeV.
\item
The radial separation between the two leading muons, denoted as $\Delta R^{\mathrm{leading, sub-leading}}_{\mu \mu}$, remains below $1.0$ and, similarly, the radial separation between the leading muon and the sub-sub-leading muon, denoted by $\Delta R^{\mathrm{leading,sub-sub-leading}}_{\mu \mu}$, must not exceed $1.2$.
\item
The difference in pseudo-rapidity between the two leading muons, $\Delta \eta^{\mathrm{leading, sub-leading}}_{\mu \mu}$, remains below $1.5$ and, similarly, the difference in pseudo-rapidity between the leading muon and the sub-sub-leading muon, denoted by $\Delta \eta^{\mathrm{leading,sub-sub-leading}}_{\mu \mu}$, must also remain below $1.5$.
\item 
The missing transverse energy, $\Et$, must exceed $150$ GeV.
\item 
The transverse momentum of the leading muon, $p^{\mu_1}_T$, should be less than $50$ GeV.
\item 
The transverse momentum of the sub-leading muon, $p^{\mu_2}_T$, must not exceed $40$ GeV.
\item 
The transverse momentum of the leading jet, $p^{j_1}_T$, must be greater than $50$ GeV.
\item
The quantity $m^{j\mu \, \Et}_{\mathrm{cluster}}$, defined in Eq.~\eqref{eq:m_cluster},  
must be grater than $200$ GeV.
\end{enumerate}

\end{itemize}


In order to provide a fair assessment of the efficacy of each selection criterion, Tab.~\ref{tab:tablecutflow_3lep} offers a comprehensive overview of the cut-flow for both signal and background events for the case of at least three muons in the final state while  Tab.~\ref{tab:tablecutflow_4lep} does so for the case of four  muons, each accompanied by some missing transverse energy. 
In each table, column 3 specifies the cross section values for both signals and backgrounds, while columns 4 through 7 provide the number of events at the 14 TeV LHC with an integrated luminosity of 3000 fb$^{-1}$ following the application of the selection criteria.

\begin{table}[h!]
\centering
\scriptsize{
\begin{tabular}{| c | c | c | c | c | c | c |}
\hline
{\bf Datasets} & {\bf Cross-section (fb)} & {\bf Pre-selection cut} & {\bf Cut-A} & {\bf Cut-B} & {\bf Cut-C} & {\bf Cut-D} \\
\hline
BP1 & $6.961$ & 17 & 16 & 16 & 13 & 13\\
\hline
BP2 & $3.733$ & 59 & 58 & 58 & 41 & 32\\
\hline
$WZ$ & $163.4068$ & 97691 & 9 & 0 & 0 & 0\\
\hline
$ZZ$  & $16.554$ & 22614 & 2 & 0 & 0 & 0\\
\hline
$WWW$ & $0.248862$ & 185 & 3 & 0 & 0 & 0\\
\hline
$WWZ/\gamma$ & $0.04978$ & 96 & 1 & 0 & 0 & 0\\
\hline
$ZZZ$ & $9.3516 \times 10^{-3}$ & 16 & 0 & 0 & 0 & 0\\
\hline
$t \bar{t} W$ & $0.606$ & 114 & 2 & 0 & 0 & 0\\
\hline
$t \bar{t} Z/\gamma$ & $0.3045$ & 136 & 1 & 0 & 0 & 0\\
\hline
$t \bar{t} W W$ & $1.279 \times 10^{-3}$ & 0 & 0 & 0 & 0 & 0 \\
\hline
$t \bar{t} t \bar{t}$ & $1.51359 \times 10^{-3}$ & 0 & 0 & 0 & 0 & 0 \\
\hline
\end{tabular}
\caption{Cross-section and number of events for signal and backgrounds after subsequent cuts at $\sqrt{s} = 14$ TeV and ${\cal L} = 3000$ fb$^{-1}$ for a final state with at-least-three-muons + $\Et$.} 
\label{tab:tablecutflow_3lep}
}
\end{table}


\begin{table}[h!]
\centering
\scriptsize{
\begin{tabular}{| c | c | c | c | c | c | c | c |}
\hline
{\bf Datasets} & {\bf Cross-section (fb)} & {\bf Pre-selection cut} & {\bf Cut-A} & {\bf Cut-B} & {\bf Cut-C} & {\bf Cut-D} \\
\hline
BP1 & $6.961$ & 2 & 1 & 1 & 1 & 1\\
\hline
BP2 & $3.733$ & 12 & 11 & 11 & 6 & 6\\
\hline
$WZ$ & $163.4068$ & 20 & 0 & 0 & 0 & 0\\
\hline
$ZZ$  & $16.554$ & 8871 & 0 & 0 & 0 & 0\\
\hline
$WWW$ & $0.248862$ & 0 & 0 & 0 & 0 & 0\\
\hline
$WWZ/\gamma$ & $0.04978$ & 41 & 0 & 0 & 0 & 0\\
\hline
$ZZZ$ & $9.3516 \times 10^{-3}$ & 6 & 0 & 0 & 0 & 0\\
\hline
$t \bar{t} W$ & $0.606$ & 1 & 0 & 0 & 0 & 0\\
\hline
$t \bar{t} Z/\gamma$ & $0.3045$ & 56 & 0 & 0 & 0 & 0 \\
\hline
$t \bar{t} W W$ & $1.279 \times 10^{-3}$ & 0 & 0 & 0 & 0 & 0\\
\hline
$t \bar{t} t \bar{t}$ & $1.51359 \times 10^{-3}$ & 136 & 0 & 0 & 0 & 0\\
\hline
\end{tabular}
\caption{Cross-section and number of events for signal and backgrounds after  subsequent cuts at $\sqrt{s} = 14$ TeV and ${\cal L} = 3000$ fb$^{-1}$ for a final state with at-least-four-muons + $\Et$.} 
\label{tab:tablecutflow_4lep}
}
\end{table}


Here, it is noteworthy that the relative size of the backgrounds amongst themselves changes drastically between the two signatures, owing to the fact that different subsets of the latter contribute to final states with different EM charges. For example, 
after imposing the pre-selection cuts to the final state with at least three muons and $\Et$, the $W^\pm Z$ process assumes prominence as a substantial background, surpassing the contribution from $ZZ$. 
In contrast, when we examine the final state with at least four muons and $\Et$, the situation is reversed. 
Furthermore, 
it is also worth noting that the higher particle multiplicity in the final state leads to cross section suppression for other contributions in the first scenario, such as $WWW/Z/\gamma$, $ZZZ$, and $t \bar{t} W/Z/\gamma$, rendering them less significant.
In contrast, for the second scenario, despite its smaller cross section, the $t \bar{t} t \bar{t}$ process emerges as a notable component due to the unique characteristics of this final state. In this context, $WWW$ does not play a significant role.

The projected significance (${\cal S}$) in the three muons plus $\Et$ channel for each BP in Tab.~\ref{tab:signi} is then calculated for the 14 TeV LHC with 3000 fb$^{-1}$ where ${\cal S}$ is defined as:
\be 
{\cal S} = \sqrt{2 \left[(S+B)\,  \log (1+\frac{S}{B}) - S\right]}\, .
\label{significance}
\ee 
Here, $S$ and $B$ denote the number of signal and background events, respectively, that have successfully passed through the sequence of selection criteria. Here, we limit ourselves to the signature with at least three muons in the final state as the one with at least four muons has no statistical relevance, owing the negligible number of surviving events. 

\begin{table}[h!]
\begin{center}
\begin{tabular}{| c | c | c |}
\hline
& $
{\cal S}$ (Pre-selection) &  ${\cal S}$ (cut-A) \\
\hline
BP1  &  0.05 $\sigma$ & 3.35 $\sigma$\\
\hline
BP2  & 0.17 $\sigma$ & 10.15 $\sigma$\\
\hline
\end{tabular}
\caption{Signal significance for the BPs at 14 TeV LHC with ${\cal L}$ = 3000 fb$^{-1}$ after pre-selection cuts and the cut-A sequence. 
}
\label{tab:signi}
\end{center}
\end{table}

Tab.~\ref{tab:signi} illustrates that, even though attaining  signal significance sufficience for evidence or discovery of a signal is challenging to start with, following the pre-selection cut, the implementation of the cut-A selection has the potential to significantly reduce background contributions while exalting the signal. Specifically, in the scenarios corresponding to BP1 and BP2, cut-A facilitates achieving signal significances exceeding $3 \, \sigma$ and $10 \, \sigma$, respectively. This is indeed the selection that we would recommend for experimental analysis. Additionally, we would like to point out the following.

\begin{itemize}
\item 
From Tab.~\ref{tab:tablecutflow_3lep}, it is evident that the application of the cut-B selection, results in the complete elimination of background events within the signal region while the number of signal events is approximately equivalent to what remains after cut-A. Therefore, cut-B may be a promising scenario for background-free searches in future High-Luminosity LHC studies.
Note that cut-C and cut-D also lead to a background-free signal, however, 
they result in a loss of some signal events.
As a result, one might consider cut-B as the most favourable criterion for our objectives.

We also note that for the at-least-three-muons + $\Et$ channel, imposing cut-A and cut-B could effectively eliminate the background interference even at Run-3 LHC with 300 fb$^{-1}$ luminosity.

\item 
Both cut-A and cut-B demonstrate remarkable efficacy in eliminating all background contributions for the at-least-four-muons + $\Et$ signal channel. Therefore, this channel holds promise for studies of BP2-type scenarios.

\item 
The trigger threshold efficiencies for muons have not been incorporated in the detector simulations thus far. Extensive testing has revealed that employing the high-level triggers designed for muons at the LHC as outlined in~\cite{CMS:2023slr} would result in the loss of approximately 30\% to 35\% of our signals when using the at-least-three-muons + $\Et$ signal channel. This loss can be attributed to the lower trigger efficiency at lower transverse momenta of the muons.

However, for the at-least-four-muons + $\Et$ signal channel, the signal loss is anticipated to be in the range of approximately 9\% to 15\%. A potential solution involves exploring the identification of muons with less stringent trigger criteria on muon transverse momentum, which has seen improvements at the LHC
This updated configuration ensures that no signals are lost in the signal channel.

\end{itemize}

\section{Conclusion and outlook}
\label{sec:conclusion}

We have investigated novel signals within the framework of a 3-Higgs Doublet Model (3HDM) at the Large Hadron Collider (LHC), where only one of the doublets acquires a Vacuum Expectation Value (VEV), maintaining a  $Z_2$ parity. The other two doublets remain \textit{inert} and do not possess a VEV, thereby forming a \textit{dark scalar sector} governed by the $Z_2$ symmetry. Within this setup, the lightest CP-even dark scalar, denoted as $H_1$, assumes the role of the Dark Matter (DM) candidate.

This scenario leads to an intriguing loop-induced decay process involving the next-to-lightest scalar, $H_2 \to H_1 \ell \bar \ell$ (where $\ell = e, \mu$), mediated by both dark CP-odd and charged scalars. This particular decay process serves as a distinct signature of the 3HDM, as it is prohibited in the 2HDM with a single inert doublet. This decay process becomes especially relevant when $H_2$ and $H_1$ exhibit close masses.

In practical terms, this signature can be observed in the cascade decay of the SM-like Higgs boson, $h\to H_1 H_2\to H_1 H_1 \ell \bar \ell$, resulting in two DM particles and di-leptons (or, possibly, di-jets). Alternatively, it can manifest as $h\to H_2 H_2\to H_1 H_1 \ell \bar \ell \ell \bar \ell$, yielding two DM particles and four-leptons (or, possibly, four-jets). The production of $h$ can occur via the gluon-gluon Fusion (ggF) channel.

However, it is important to note that the di-lepton signal competes with the tree-level channel $q\bar q\to H_1H_1Z^*\to H_1H_1 \ell \bar \ell$, which can be overwhelming. Therefore, we have concentrated here on the $h\to H_2H_2 \to H_1 H_1 \ell \bar \ell \ell \bar \ell$ case only and targeted two possible signatures, one involving at least three muons and and another at least four muons, which we have preferred to electrons as the former (unlike the latter) allow one to target small transverse momentum regions, which are those pertaining to our signals, as the mass difference between the $H_2$ and $H_1$ states is taken to be small.

In  order to explore this scenario, we have established BPs that align with collider, DM and cosmological data.
We delve into the interplay between these modes, demonstrating that the resulting detector signature, characterized by $\Et \ell \bar \ell$ or $\Et \ell \bar \ell \ell \bar \ell$, with an invariant mass of $ \ell \bar \ell$ significantly smaller than $m_Z$, could potentially be detected at the High-Luminosity (HL) phase of the LHC.

We present our results for signals with at least three muons + $\Et$ or  at least four muons + $\Et$  in the final state after applying certain cuts and show that both final states have signal significance sufficient for discovery at HL-LHC. We also show that the at-least-three-muons + $\Et$ signal has the ponetial to be porbed at 14 TeV LHC with 300 fb$^{-1}$ for BP2 scenario.

\subsection*{Acknowledgements}
SM acknowledges support from the STFC Consolidated Grant ST/L000296/1 and is partially financed through the NExT Institute.
VK and AD acknowledge financial support from the Science Foundation Ireland Grant 21/PATHS/9475 (MOREHIGGS) under the SFI-IRC Pathway Programme.

\appendix

\section{Loop-level decays of $H_2 \to H_1 \gamma^*$}
\label{app:loop-decay-diagrams}

The corresponding loops go through triangle and bubble diagrams with $H^\pm_i$ and $W^\pm$ entering, as shown in Figs.~\ref{fig:triangle-decays}-\ref{fig:bubble-decays}.
Note that there are also box diagrams which contribute to the process $H_2 \to H_1 \ell  \bar\ell  $, presented in Fig.~\ref{fig:box}. Here, the $\ell  \bar\ell  $ pair is produced through the SM gauge-lepton tree-level vertices, without producing an intermediate off-shell photon. The corresponding topologies also see the contribution of inert, both charged and neutral pseudo-scalars.
However, due to the mass suppression, the contribution from the box diagrams is small, of order 10\%, and it leaves the results practically unaffected. Therefore, we do not show the results of these box diagrams in the numerical scans. 

Let us emphasise that one could attempt constructing analogous diagrams to those in Figs.~\ref{fig:triangle-decays}-\ref{fig:bubble-decays} with $H_2$ replaced by $A_1$ or $A_2$, leading to 
$A_i \to H_1 \gamma^*, i=1,2$. However, this decay would lead to a CPV process, while the model we analyse here is explicitly CPC. 
Note also that spin conservation requires that it is only the scalar polarization of the virtual photon that contributes to the
$H_2\to H_1\gamma^*$ transition.
To check the validity of the calculations, we have explicitly verified
this to be the case, as there are cancellations between diagrams that lead to the amplitude being equal to zero otherwise, as discussed in detail in a previous publication in \cite{Cordero:2017owj}.
Also note that the process $A_i\to H_1 Z^*$ does exist at tree-level in both the I(2+1)HDM (for $i=1,2$)  and  I(1+1)HDM (for $i=1$) and contributes to the $\Et  \ell  \bar\ell  $ signature, as discussed previously. However, in the interesting regions of the parameter space where the invariant mass of the $\ell  \bar\ell  $ pair is small, i.e., $\ll m_Z$, this process is sub-dominant.

\begin{minipage}{\linewidth}
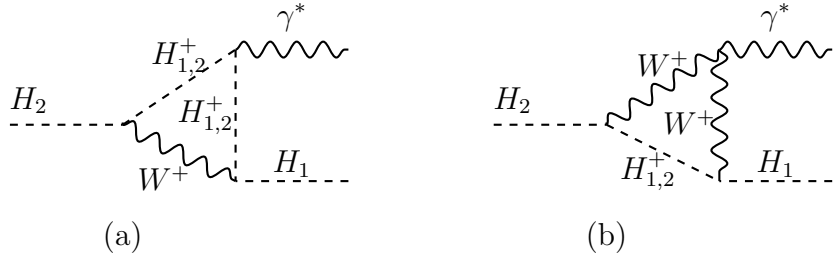
\begin{figure}[H]
\hspace{1.5cm}
\begin{tikzpicture}[thick,scale=1.0]
\draw (1.5,0) -- node[black,above,xshift=-0.1cm,yshift=0.0cm] {$ $} (1.5,0.03);
\draw[dashed] (0,0) -- node[black,above,xshift=-0.5cm,yshift=0cm] {$H_{2}$} (1.5,0);
\draw[dashed] (1.5,0) -- node[black,above,xshift=0cm,yshift=0cm] {$H^+_{1,2}$} (3,1.0);
\draw[dashed] (4.5,-0.75) -- node[black,above,yshift=-0.1cm,xshift=0cm] {$H_1$} (3,-0.75);
\draw[decorate,decoration={snake,amplitude=3pt,segment length=10pt}](1.5,0) -- node[black,above,yshift=-0.65cm,xshift=-0.2cm]  {$W^+$} (3,-0.75);
\draw[dashed](3,1) -- node[black,above,yshift=-0.4cm,xshift=-0.4cm]  {$H^+_{1,2}$} (3,-0.75);
\draw[decorate,decoration={snake,amplitude=3pt,segment length=10pt}](3,1) -- node[black,above,yshift=0.1cm,xshift=0cm]  {$\gamma^*$} (4.5,1);
\node at (1.5,-1.5) {(a)};
\end{tikzpicture}
\hspace{1.5cm}
\begin{tikzpicture}[thick,scale=1.0]
\draw (1.5,0) -- node[black,above,xshift=-0.1cm,yshift=0.0cm] {$ $} (1.5,0.03);
\draw[dashed] (0,0) -- node[black,above,xshift=-0.5cm,yshift=0cm] {$H_{2}$} (1.5,0);
\draw[decorate,decoration={snake,amplitude=3pt,segment length=10pt}] (1.5,0) -- node[black,above,xshift=0cm,yshift=0cm] {$W^+$} (3,1.0);
\draw[dashed] (4.5,-0.75) -- node[black,above,yshift=-0.1cm,xshift=0cm] {$H_1$} (3,-0.75);
\draw[dashed](1.5,0) -- node[black,above,yshift=-0.65cm,xshift=-0.2cm]  {$H^+_{1,2}$} (3,-0.75);
\draw[decorate,decoration={snake,amplitude=3pt,segment length=10pt}](3,1) -- node[black,above,yshift=-0.4cm,xshift=-0.4cm]  {$W^+$} (3,-0.75);
\draw[decorate,decoration={snake,amplitude=3pt,segment length=10pt}](3,1) -- node[black,above,yshift=0.1cm,xshift=0cm]  {$\gamma^*$} (4.5,1);
\node at (1.5,-1.5) {(b)};
\end{tikzpicture}
\vspace{0.5cm}
\caption{Triangle diagrams contributing to the $H_2 \to H_1 \gamma^*$ decay, where the lightest inert is absolutely stable and hence invisible,
while $\gamma^*$ is a virtual photon that couples to the $\ell \bar{\ell}$ pair.  Analogous diagrams cannot be constructed if the initial particle is $A_{1}$ or $A_2$.}
\label{fig:triangle-decays} 
\end{figure}    
\end{minipage}

\begin{minipage}{\linewidth}
\begin{figure}[H]
\begin{tikzpicture}[thick,scale=1.0]
\draw[dashed] (0,0) -- node[black,above,sloped,yshift=-0.1cm,xshift=-0.4cm] {$H_{2}$} (1,0);
\draw[dashed] (3,0) -- node[black,above,yshift=-0.3cm,xshift=0.5cm] {$H_1$} (4.2,-1.1);
\draw[decorate,decoration={snake,amplitude=3pt,segment length=10pt}] (3,0) -- node[black,above,yshift=-0.4cm,xshift=0.5cm] {$\gamma^*$} (4.2,1.1);
\draw[dashed]  (1,0) node[black,above,sloped,yshift=0.95cm,xshift=1.05cm] {$H^+_{1,2}$}  arc (180:0:1cm) ;
\draw[decorate,decoration={snake,amplitude=3pt,segment length=10pt}]  (1,0) node[black,above,sloped,yshift=-0.95cm,xshift=1.05cm] {$W^+$}  arc (-180:0:1cm) ;
\node at (1.5,-2.3) {(A)};
\end{tikzpicture}
\hspace{2mm}
\begin{tikzpicture}[thick,scale=1.0]
\draw[dashed] (0,0) -- node[black,above,xshift=-0.3cm,yshift=-0.1cm] {$H_{2}$} (1.5,0);
\draw[dashed] (1.5,0) -- node[black,above,yshift=0.4cm,xshift=0.2cm] {$H_1$} (2.7,1.4);
\draw[decorate,decoration={snake,amplitude=3pt,segment length=10pt}] (2.8,-1.2) -- node[black,above,yshift=-0.3cm,xshift=0.9cm] {$\gamma^*$} (4.3,-1.7);
\draw[dashed]  (1.5,0) node[black,above,sloped,yshift=0.2cm,xshift=1.3cm] {$H^+_{1,2}$}  arc (140:-40:0.9cm) ;
\draw[dashed]  (1.5,0) node[black,above,sloped,yshift=-1.4cm,xshift=0.55cm] {$H^+_{1,2}$}  arc (-220:-40:0.9cm) ;
\node at (1.5,-2.3) {(B)};
\end{tikzpicture}
\begin{tikzpicture}[thick,scale=1.0]
\draw[dashed] (0,0) -- node[black,above,xshift=-0.3cm,yshift=-0.1cm] {$H_{2}$} (1.5,0);
\draw[decorate,decoration={snake,amplitude=3pt,segment length=10pt}] (1.5,0) -- node[black,above,yshift=0.4cm,xshift=0.2cm] {$\gamma^*$} (2.7,1.4);
\draw[dashed] (2.8,-1.2) -- node[black,above,yshift=-0.3cm,xshift=0.9cm] {$H_1$} (4.3,-1.7);
\draw[dashed]  (1.5,0) node[black,above,sloped,yshift=0.2cm,xshift=1.3cm] {$H^+_{1,2}$}  arc (140:-40:0.9cm) ;
\draw[decorate,decoration={snake,amplitude=3pt,segment length=10pt}]  (1.5,0) node[black,above,sloped,yshift=-1.4cm,xshift=0.55cm] {$W^+$}  arc (-220:-40:0.9cm) ;
\node at (1.5,-2.3) {(c)};
\end{tikzpicture}
\vspace{0.5cm}
\caption{Bubble  diagrams contributing to the $H_2 \to H_1 \gamma^*$ decay, where the lightest inert particle is absolutely stable and hence invisible,
while $\gamma^*$ is a virtual photon that couples to the $\ell \bar{\ell}$ pair.  Analogous diagrams cannot be constructed if the initial particle is $A_{1}$ or $A_2$.}
\label{fig:bubble-decays} 
\end{figure}
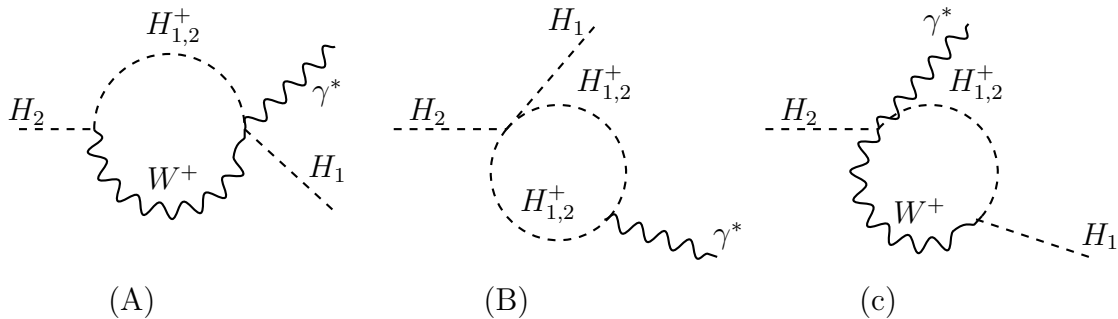    
\end{minipage}

\begin{minipage}{\linewidth}
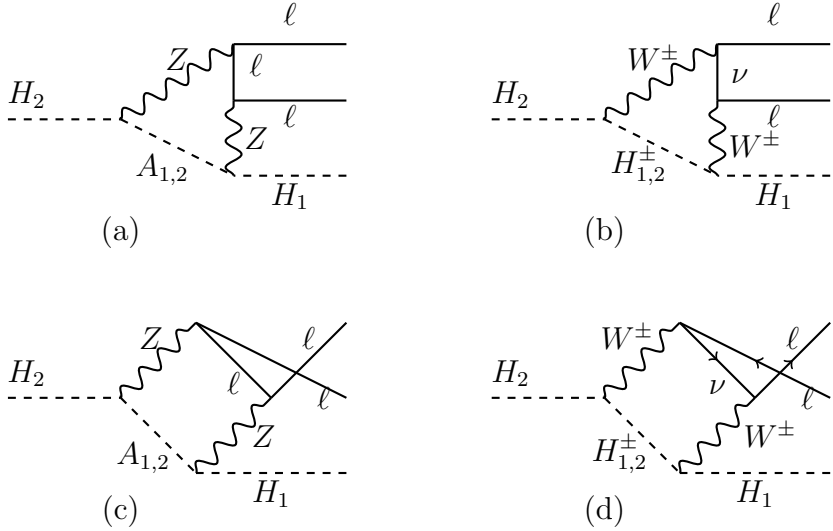
\begin{figure}[H]
\hspace{1.5cm}
\begin{tikzpicture}[thick,scale=1.0]
\draw (1.5,0) -- node[black,above,xshift=-0.1cm,yshift=0.0cm] {$ $} (1.5,0.03);
\draw[dashed] (0,0) -- node[black,above,xshift=-0.5cm,yshift=0cm] {$H_{2}$} (1.5,0);
\draw[decorate,decoration={snake,amplitude=3pt,segment length=10pt}] (1.5,0) -- node[black,above,xshift=0cm,yshift=0cm] {$Z$} (3,1.0);
\draw[dashed] (4.5,-0.75) -- node[black,above,yshift=-0.6cm,xshift=0cm] {$H_1$} (3,-0.75);
\draw[dashed](1.5,0) -- node[black,above,yshift=-0.65cm,xshift=-0.2cm]  {$A_{1,2}$} (3,-0.75);
\draw[solid](3,1) -- node[black,above,yshift=0.1cm,xshift=0cm]  {$\ell$} (4.5,1);
\draw[solid](3,0.25) -- node[black,above,yshift=-0.5cm,xshift=0cm]  {$\ell$} (4.5,0.25);
\draw[solid](3,0.25) -- node[black,above,yshift=-0.2cm,xshift=0.3cm]  {$\ell$} (3,1);
\draw[decorate,decoration={snake,amplitude=3pt,segment length=10pt}] (3,-0.75) -- node[black,above,yshift=-0.3cm,xshift=0.3cm] {$Z$} (3,0.25);
\node at (1.5,-1.5) {(a)};
\end{tikzpicture}
\hspace{1.5cm}
\begin{tikzpicture}[thick,scale=1.0]
\draw (1.5,0) -- node[black,above,xshift=-0.1cm,yshift=0.0cm] {$ $} (1.5,0.03);
\draw[dashed] (0,0) -- node[black,above,xshift=-0.5cm,yshift=0cm] {$H_{2}$} (1.5,0);
\draw[decorate,decoration={snake,amplitude=3pt,segment length=10pt}] (1.5,0) -- node[black,above,xshift=-0.1cm,yshift=0cm] {$W^\pm$} (3,1.0);
\draw[dashed] (4.5,-0.75) -- node[black,above,yshift=-0.6cm,xshift=0cm] {$H_1$} (3,-0.75);
\draw[dashed](1.5,0) -- node[black,above,yshift=-0.65cm,xshift=-0.3cm]  {$H^\pm_{1,2}$} (3,-0.75);
\draw[solid](3,1) -- node[black,above,yshift=0.1cm,xshift=0cm]  {$\ell$} (4.5,1);
\draw[solid](3,0.25) -- node[black,above,yshift=-0.5cm,xshift=0cm]  {$\ell$} (4.5,0.25);
\draw[solid](3,0.25) -- node[black,above,yshift=-0.3cm,xshift=0.3cm]  {$\nu$} (3,1);
\draw[decorate,decoration={snake,amplitude=3pt,segment length=10pt}] (3,-0.75) -- node[black,above,yshift=-0.4cm,xshift=0.5cm] {$W^\pm$} (3,0.25);
\node at (1.5,-1.5) {(b)};
\end{tikzpicture}\\[2mm]

\hspace{1.5cm}
\begin{tikzpicture}[thick,scale=1.0]
\draw (1.5,0) -- node[black,above,xshift=-0.1cm,yshift=0.0cm] {$ $} (1.5,0.03);
\draw[dashed] (0,0) -- node[black,above,xshift=-0.5cm,yshift=0cm] {$H_{2}$} (1.5,0);
\draw[decorate,decoration={snake,amplitude=3pt,segment length=10pt}] (1.5,0) -- node[black,above,xshift=-0.1cm,yshift=0cm] {$Z$} (2.5,1.0);
\draw[dashed](1.5,0) -- node[black,above,yshift=-0.65cm,xshift=-0.2cm]  {$A_{1,2}$} (2.5,-1);
\draw[decorate,decoration={snake,amplitude=3pt,segment length=10pt}] (2.5,-1) -- node[black,above,yshift=-0.3cm,xshift=0.4cm] {$Z$} (3.5,0.);
\draw[solid](2.5,1) -- node[black,above,yshift=-0.6cm,xshift=0cm]  {$\ell$} (3.5,0);
\draw[dashed] (4.5,-1) -- node[black,above,yshift=-0.6cm,xshift=0cm] {$H_1$} (2.5,-1);
\draw[solid](3.5,0) -- node[black,above,yshift=0.0cm,xshift=0cm]  {$\ell$} (4.5,1);
\draw[solid](4.5,0) -- node[black,above,yshift=-0.8cm,xshift=0.7cm]  {$\ell$} (2.5,1);
\node at (1.5,-1.5) {(c)};
\end{tikzpicture}
\hspace{1.5cm}
\begin{tikzpicture}[thick,scale=1.0]
\draw (1.5,0) -- node[black,above,xshift=-0.1cm,yshift=0.0cm] {$ $} (1.5,0.03);
\draw[dashed] (0,0) -- node[black,above,xshift=-0.5cm,yshift=0cm] {$H_{2}$} (1.5,0);
\draw[decorate,decoration={snake,amplitude=3pt,segment length=10pt}] (1.5,0) -- node[black,above,xshift=-0.2cm,yshift=0cm] {$W^\pm$} (2.5,1.0);
\draw[dashed](1.5,0) -- node[black,above,yshift=-0.65cm,xshift=-0.3cm]  {$H^\pm_{1,2}$} (2.5,-1);
\draw[decorate,decoration={snake,amplitude=3pt,segment length=10pt}] (2.5,-1) -- node[black,above,yshift=-0.3cm,xshift=0.7cm] {$W^\pm$} (3.5,0.);
\draw[particle](2.5,1) -- node[black,above,yshift=-0.6cm,xshift=0cm]  {$\nu$} (3.5,0);
\draw[dashed] (4.5,-1) -- node[black,above,yshift=-0.6cm,xshift=0cm] {$H_1$} (2.5,-1);
\draw[particle](3.5,0) -- node[black,above,yshift=0.0cm,xshift=0cm]  {$\ell$} (4.5,1);
\draw[particle](4.5,0) -- node[black,above,yshift=-0.8cm,xshift=0.7cm]  {$\ell$} (2.5,1);
\node at (1.5,-1.5) {(d)};
\end{tikzpicture}
\caption{Box diagrams contributing to $H_2 \to H_1 \ell \ell$. }
\label{fig:box} 
\end{figure}    
\end{minipage}


\bibliographystyle{OurBibTeX}

\end{document}